\documentclass[aps,prl,twocolumn,amsmath,amssymb,superscriptaddress,longbibliography]{revtex4-2}
\usepackage[breaklinks,colorlinks,bookmarks=false,citecolor=blue,linkcolor=red,urlcolor=blue]{hyperref}
\usepackage{blindtext,enumitem,graphicx,dcolumn,color,xcolor,amsmath,braket,bm,changes,chemformula,cleveref,times}

\graphicspath{{figures/}}
\usepackage{blindtext}
\usepackage{changes}
\usepackage{subcaption}
\usepackage[utf8]{inputenc}
\usepackage[english]{fixme}
\usepackage{wasysym}            
\crefname{equation}{Eq.}{Eqs.}
\Crefname{equation}{Eq.}{Eqs.}
\crefname{figure}{Fig.}{Figs.}
\Crefname{figure}{Fig.}{Figs.}
\crefname{section}{Sec.}{Secs.}
\Crefname{section}{Sec.}{Secs.}

\usepackage{caption}
\begin{document}
\title{The renormalized classical spin liquid on the ruby lattice}

\author{Zhenjiu Wang}
\email{zhenjiu.wang@physik.uni-muenchen.de}
\affiliation{Department of Physics and Arnold Sommerfeld Center for Theoretical Physics (ASC), Ludwig-Maximilians-Universit\"at M\"unchen, Theresienstr. 37, M\"unchen D-80333, Germany}
\affiliation{Munich Center for Quantum Science and Technology (MCQST), Schellingstr. 4, D-80799 M\"unchen, Germany}
\author{Lode Pollet}
\affiliation{Department of Physics and Arnold Sommerfeld Center for Theoretical Physics (ASC), Ludwig-Maximilians-Universit\"at M\"unchen, Theresienstr. 37, M\"unchen D-80333, Germany}
\affiliation{Munich Center for Quantum Science and Technology (MCQST), Schellingstr. 4, D-80799 M\"unchen, Germany}

\begin{abstract}
The recent experimental detection of the onset of a dynamically prepared, gapped $Z_2$ quantum spin liquid on the ruby lattice brought the physics of frustrated magnetism and lattice gauge theory to Rydberg tweezer arrays~\cite{semeghini_probing_2021}.  The thermodynamic properties of such models remain inadequately addressed, yet knowledge thereof is indispensable if one wants to prepare large, robust, and long-lived quantum spin liquids. Using large scale quantum Monte Carlo simulations we find in the {\it PXP} model a renormalized classical spin liquid with constant entropy density $S/N$ approaching $\ln(2)/6$ in the thermodynamic limit for all moderate and large values of the detuning $\delta$ and starting from $T/\Omega \sim 0.5$ (in units of the Rabi frequency $\Omega$) down  to the lowest temperatures we could simulate, $T/\Omega \sim 0.01$. With Van der Waals interactions, constant entropy plateaus are still found but its value shifts with $\delta$. We comment the adiabatic approximation to the dynamical ramps for the electric degrees of freedom.
\end{abstract}

\date{\today}

\maketitle

{\it Introduction -- }  There is considerable interest in realizing $Z_2$ quantum spin liquids (QSL)~\cite{savary_quantum_2017}, which are special states of gapped matter with long-range entanglement~\cite{kitaev_2003}, topological order~\cite{wen_colloquium_2017}, and fractionalized excitations~\cite{wilczek_fractional_1982}. Techniques to protect quantum information with such systems are actively pursued~\cite{nayak_rmp_2008, terhal_rmp_2015,satzinger_toric_2021}. They also establish a clean connection between QSLs and lattice gauge theory~\cite{wegner_1971,fradkin_phase_1979,kitaev_2003}, where the QSL is the deconfined phase of matter whose excitations (eg, anyons) can be separated at no energy cost.
Their realization has so far been elusive in microscopic spin or bosonic models. 

Theoretically, the existence of such states is well established in exactly soluble models such as the toric code~\cite{kitaev_2003}, Kitaev's spin liquid model on the honeycomb lattice~\cite{kitaev_honeycomb_2006}, and quantum dimer models~\cite{rokhsar_superconductivity_1988} on nonbipartite lattices~\cite{sachdev_1992,moessner_quantum_2008}, such as the triangular~\cite{moessner_2001} and kagome lattice~\cite{misguich_2002}.
In other cases they have been established numerically, for instance, the $S=1$  Kitaev model~\cite{Baskaran_2008,Rousochatzakis2018,chen_spinone_kitaev_2022}, and recently in the {\it PXP} (or blockade) model on the ruby lattice~\cite{verresen_prediction_2021}. The latter study motivated an experiment at Harvard in which Rydberg atoms in tweezer arrays were quasi-adiabatically prepared and in which typical signatures of a QSL have been detected~\cite{semeghini_probing_2021}. This was only possible thanks to the spectacular recent experimental advances in Rydberg atom platforms~\cite{bernien_probing_2017,browaeys_many-body_2020,ebadi_quantum_2021,ebadi_optimization_2022}. 

The possibility of studying QSLs in a Rydberg context brings additional advantages compared to traditional condensed matter approaches: more control over interactions and densities, in-situ detection techniques such as equal-time string correlators, little to no disorder and special preparation techniques to name just a few. A major difference is however the absence of a bath and thus of cooling possibilities. Candidate condensed matter materials typically show a temperature range in which a classical spin liquid (CSL) is seen~\cite{KnolleMoessner_review}, which consists of an exponential number of states satisfying (typically) a local constraint and defying any type of magnetic order. The QSL, requiring much lower temperatures, originates from a quantum term coupling states in this macroscopically degenerate manifold. In the Rydberg system, the same mechanism is at work, although nearly all theoretical studies focused on either the ground state or the quantum evolution of a single quantum state, showing features of a QSL.

Nevertheless, large, robust and long-lived QSLs can only be found in a dynamical experiment if the ground state has  a stable QSL. To mitigate unavoidable experimental imperfections (finite duration of the time-dependent protocol, spontaneous emission, black-body radiation, dephasing, state preparation and measurement errors, jitter, etc) one also desires a large gap, which is challenging because the gauge-theoretic description is only {\it emergent}. 
In this Letter, we perform large-scale quantum Monte Carlo simulations on the ruby lattice (cf. Ref.~\cite{patil2023quantum}) to give a more traditional condensed matter perspective, finding a renormalized CSL over a very wide temperature range of direct relevance to experiment. We comment on the use of the proposed~\cite{verresen_prediction_2021} equal-time parity string correlators. We utilize local updates as well as the cluster algorithm devised by Alet {\it et al}~\cite{Alet2006}  which significantly enhances the ergodicity of the simulations at large detunings.\\

\begin{figure}
    \centering
    \includegraphics[width=0.549\textwidth]{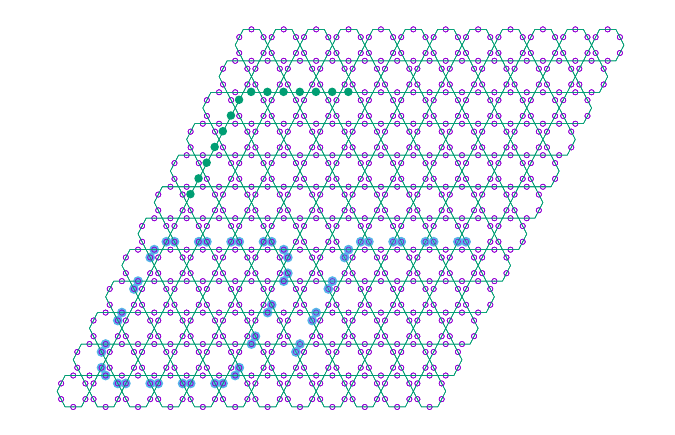}   
    \caption{Geometrical structure of Rydberg atoms (purple circles) on the Ruby lattice and illustrations of the closed loop $l$ for $\mathcal{Z}(l)$ (bottom left, Eq.~\ref{Eq:def_WC}),  and  the  open line segments for the diagonal $C_m(l)$ (bottom right, Eq.~\ref{eq:Cm}) and off-diagonal $C_e(l)$ (top left, Eq.~\ref{Eq:def_X}).  
    }
    \label{fig:Lattice}
\end{figure}

{\it Model --} We consider the following Hamiltonian describing Rydberg atoms on a two-dimensional ruby lattice,
\begin{equation}
\begin{aligned}
  H^{Ryd} & = - \frac{\Omega}{2} \sum_j ( b_j + b^{\dagger}_j )  
   -  \delta \sum_j n_j     
   + \sum_{  i < j  } V_{ij} n_i n_j,  
 \end{aligned}   
\end{equation}    
where $n_i$ denotes the occupation number of a Rydberg excitation at site $i$, $ V_{ ij } $ is the interaction between Rydberg excitations at site $i$ and $j$,  $ \delta $ is the detuning and $ \Omega $ the Rabi frequency. We set $\Omega =1$ as the energy unit and the shortest interparticle distance  $a=1$ as the length unit.  Simulations are typically done with periodic boundary conditions for $L \times L$ unit cells for the geometry shown in Fig.~\ref{fig:Lattice} with $L=12$ unless otherwise written. The total number of sites is $N = 6L^2$.

Density matrix renormalization group  (DMRG) studies with full Van der Waals (VdW) interactions $ V_{i j} = \Omega ( R_b / | \bm{i} - \bm{j} |)^6 $ and blockade radius $R_b = 2.4a$ found no thermodynamically stable quantum spin liquid~\cite{semeghini_probing_2021,sahay_quantum_2023}.  We focus therefore first on the simpler case of a blockade or {\it PXP}-model with 
\begin{equation}
   \begin{aligned} 
   V(r) \equiv & \, \infty   \quad \  \text{if} \quad r \leq 2 a \\ 
     & 0  \qquad  \text{if} \quad r > 2a,
   \end{aligned}
\end{equation} 
where DMRG studies found the following ground state phase diagram~\cite{verresen_prediction_2021}:
(i) A trivial phase for $\delta \lesssim 1.45$ where the potential energy is irrelevant and the ground state translationally invariant; (ii) A valence bond solid (VBS) with large unit cell for $ \delta \gtrsim 2.1$ where  lattice and rotational symmetry are spontaneously broken; and (iii) in between  a $Z_2$ gapped QSL. In the language of lattice gauge theory, these phases are expected based on the work by Fradkin and Shenker~\cite{fradkin_phase_1979}, corresponding to the Higgs phase ($e$-condensate), confined phase ($m$-condensate), and topologically deconfined phase, respectively. 

{\it Higgs transition -- } In the trivial phase, the quantum term in $\Omega$ dominates. The ground state is then connected to a product state in which each spin points along $x$.  Upon increasing the detuning $\delta$ we expect a quantum phase transition between the trivial and the QSL phase, which is continuous. In the absence of a (simple) local order parameter, the (intensive) fidelity susceptibility provides a direct way of detecting the quantum phase transition. Following Refs.~\cite{you_fidelity_2007,schwandt_quantum_2009,albuquerque_quantum_2010,wang_fidelity_2015} its extension to finite temperature can be expressed as
\begin{equation}
   \chi_F = \frac{1}{N} \int_0^{\beta/2} \tau  \left( \left< H_1(\tau) H_1(0) \right> - \left< H_1 \right>^2 \right) d\tau,
\end{equation}
where $H_1(\tau)$ is the Heisenberg operator of the perturbation driving the transitions, and which can directly be evaluated in  quantum Monte Carlo simulations because of the diagonal nature of the perturbation, $H_1(\tau) = n(\tau)$, in the computational Fock basis.

\begin{figure}
    \centering
    \includegraphics[width=\columnwidth]{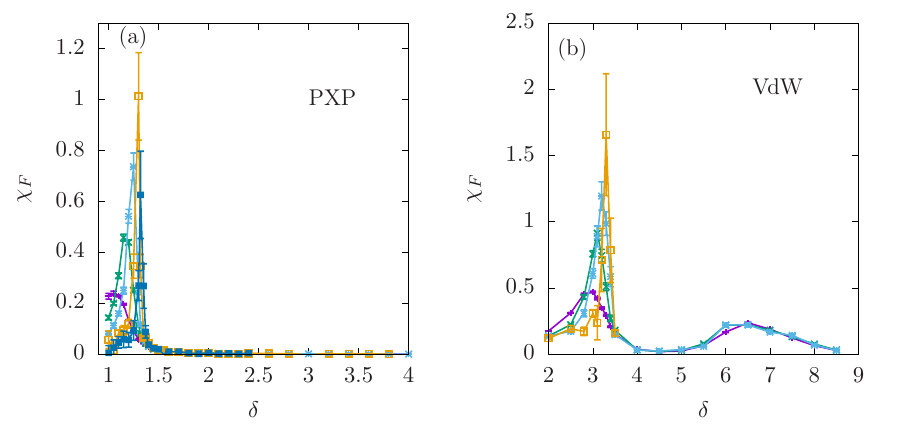}
    \caption{ 
    (a) Fidelity susceptibility $\chi_F$ for $\beta=10$, $15$, $20$, $30$ and $40$ in the {\it PXP} model, identifying the transition from trivial to spin liquid but failing to capture a second transition expected for large $\delta$ in the ground state.
    (b) Same in the VdW model, identifying the transition from trivial to spin liquid and a broad crossover to a trivial paramagnet around $\delta=6.5$.
    }\label{fig:fidelity}
\end{figure}

As shown in Fig.~\ref{fig:fidelity}(a),  for any finite system size $\chi_F$ shows a peak whose location can be extrapolated as a function of $1/\beta$ to yield the quantum critical point,  $\delta \approx 1.45$, in line with Refs.~\cite{verresen_prediction_2021,semeghini_probing_2021,tarabunga_gauge-theoretic_2022}. Equally striking is the absence of a second bump for the QSL-VBS transition, even for the lowest temperatures.

\begin{figure}
    \centering
    \includegraphics[width=\columnwidth]{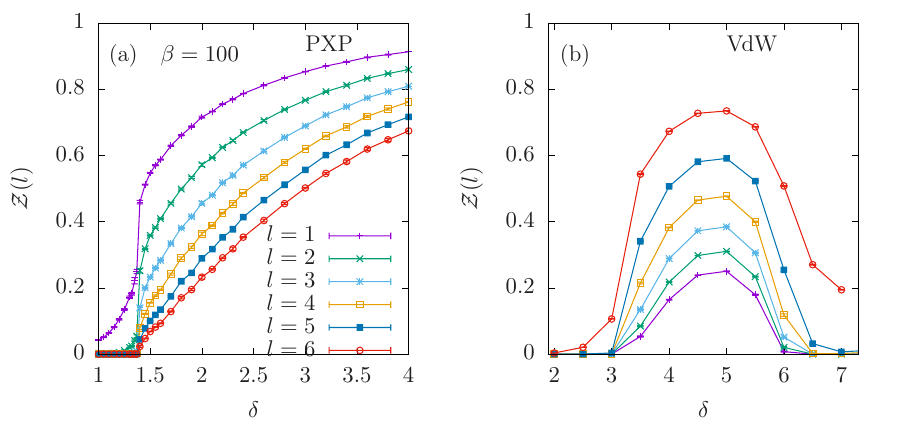}    
    \caption{
   (a) The diagonal closed-loop operator $\mathcal{Z}(l)$ as a function of $\delta$ probing Gauss law for lengths $l=1$ to $6$ at $\beta=100$ in the {\it PXP} model. (b) Same at  $\beta = 20$  in the VdW model. The results are consistent with Fig.~\ref{fig:fidelity}.
    }\label{fig:Loop}
\end{figure}

A first hint of the nature of this phase can be inferred from the absolute value of the diagonal loop operator (cf the lower left part of Fig.~\ref{fig:Lattice}),
\begin{equation}\label{Eq:def_WC}
   \mathcal{Z}(l) \equiv  \left|  \left< \prod_{ i \in C } (-1)^{n_i}  \right> \right|.
\end{equation}   
It counts the number of enclosed vertices, $\mathcal{Z}(l) = \vert (-1)^{\rm enclosed \, vertices} \vert$, cf. Refs~\cite{verresen_prediction_2021, semeghini_probing_2021}. For a perfect dimer covering $\mathcal{Z}(l) = 1$, but the presence of  monomers acting as $e-$anyons reduces the magnitude. 
As is seen in Fig.~\ref{fig:Loop}(a), for a very low temperature $\beta =100$, $\mathcal{Z}(l)$ also locates the transition from a trivial to a spin liquid at $\delta \approx 1.45$. 


%

\begin{figure}
    \centering
    \includegraphics[width=\columnwidth]{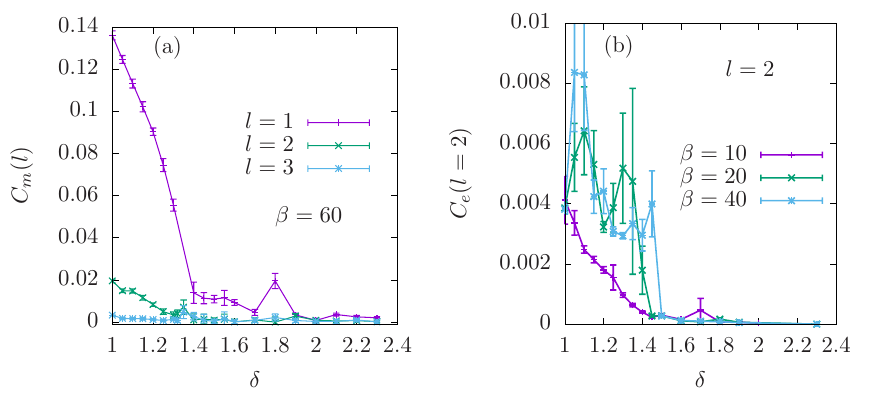}   
    \caption{ 
    (a) The quantity $C_m(l)$ as a function of $\delta$  for lengths $l=1$, $2$ and $3$ at $\beta=60$, failing to find the VBS phase.
    (b) The quantity $C_e(l)$ as a function of $\delta$  for length $l=2$, at $\beta=10, 20$ and $40$, computed in a discrete time word line approach, indicative of the trivial to spin liquid phase. Calculations are done for the {\it PXP} model.
    }\label{fig:BFFM}
\end{figure}

\begin{figure}
    \centering
    \includegraphics[width=\columnwidth]{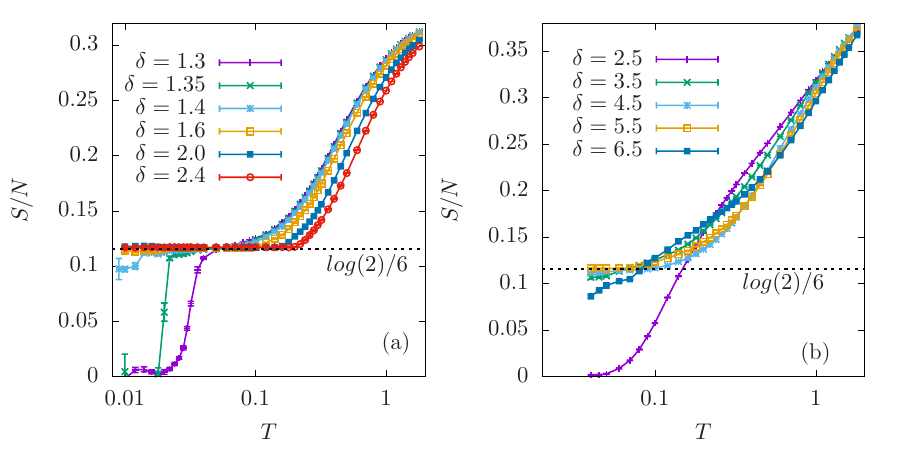}   
    \caption{  
    Entropy density as a function of temperature in the {\it PXP} (a) and VdW (b) models for various values of $\delta$.  
        }\label{fig:entropy}
\end{figure}

{\it String order parameters -- } Intriguingly, Ref.~\cite{verresen_prediction_2021} put forward a pair of experimentally measurable string operators, which are equal time variants of the order parameters suggested by Bricmont, Fr\"ohlich, Marcu and Fredenhagen~\cite{bricmont_order_1983, fredenhagen_confinement_1986,fredenhagen_dual_1988, gregor_diagnosing_2011}, namely the diagonal and off-diagonal parity string operators, which can be shown to be related to the Wilson and the 't Hooft lines~\cite{tarabunga_gauge-theoretic_2022}. Colloquially speaking, they probe anti-deconfinement. The diagonal string correlator is non-zero in the VBS phase, whereas the off-diagonal one is non-zero in the trivial phase~\cite{verresen_prediction_2021}. The deconfined phase is then found where both are zero (while the closed loop operators scale with a perimeter law).
Because the closed loop operator is negligibly small in the trivial phase ( cf. Fig.~\ref{fig:Loop}), we analyzed  the absolute value of the numerator of the diagonal component of Fredenhagen-Marcu (FM) order parameter {equation}\label{Eq:def_Z} (cf. the diagonal parity string correlator)
\begin{equation}  \label{eq:Cm}
   C_{m} (l) \equiv \left| \langle \prod_{ i \in \Gamma_Z(l) } (-1)^{n_i}  \rangle \right|, 
\end{equation}
where path $\Gamma_Z (l)$ with length $l$ is illustrated in Fig.~\ref{fig:Lattice}. The full FM order parameter normalizes $C_m(l)$ by dividing it through the square root of the corresponding closed loop parity, as shown in Fig.~\ref{fig:Lattice}.

Similarly, we analyzed the numerator of the off-diagonal component of the BFFM order parameter which  can capture the transition to the trivial phase,
\begin{equation}\label{Eq:def_X}
   C_{e} (l) \equiv \langle \prod_{ i \in \Gamma_X(l) } ( b^{\dagger}_i + b_i   )   \rangle,  
\end{equation} 
where $\Gamma_X (l)$  is again shown in Fig.~\ref{fig:Lattice}. This poor man's formula deviates from the exact off-diagonal BFFM order parameter introduced in Refs.~\cite{verresen_prediction_2021, semeghini_probing_2021} for technical reasons. Since it creates two electric charges, it plays a similar role and suffices for our purposes.


We readily see in Fig.~\ref{fig:BFFM}(a) that $C_m (l) $  does not exhibit any finite value for $\delta>2$, even for short lengths $l$. The failure in observing spontaneous lattice translational symmetry breaking for large $\delta$  is consistent with the absence of a singularity in $\chi_F$. We note in passing that the failure to detect the VBS phase experimentally~\cite{semeghini_probing_2021} is in line with our observations, and expected (see below).  Meanwhile, as shown in Fig.~\ref{fig:BFFM}(b), the  electric particle condensation transition from the deconfined  to the paramagnetic state is well captured by $ C_e(l)$ around $\delta \approx 1.45$ for $l=2$. As this transition is well understood, we see no merit in going to larger values of $l$, which is technically challenging in the QMC. The value on the spin liquid side of the shown $ C_e(l)$ is negligibly small~\footnote{From the arguments we present below, the off-diagonal closed loop operator is also expected to be vanishingly small}.   We thus conclude that the QMC finds a single phase for $\delta \gtrsim 1.45$ down to at least $T \approx 0.01$ (and much lower temperatures for greater values of $\delta$, see below).  This phase is, however, a renormalized classical spin liquid, while the relevance of the (off-diagonal) BFFM string orders applies primarily to a QSL.


{\it Entropic considerations --}  One can resolve a CSL by witnessing a plateau in the entropy at non-zero value over a wider range of temperatures, reflecting the exponential number of states satisfying the local constraint of the CSL. The entropy can be computed through an energy integration as 
\begin{equation}
  S(T) = S( \infty ) - \int_{ T }^{ \infty } \frac{ E(T') - E(T ) }{ T'^2 } d T',
\end{equation}
%
%
%

As shown in Fig.~\ref{fig:entropy}(a),  such a plateau extending from $T \sim 0.5$ down to (at least) $T \sim 0.01$ is found for all values of the detuning $\delta$ not belonging to the trivial phase. Remarkably, the plateau acquires a constant value, $S/N = \log(2)/6$ in the thermodynamic limit, independent of $\delta$, but the onset temperature and energy of the CSL are $\delta$-dependent. Close to the transition, for $\delta$ in the range  $1.3 - 1.4$ we see a drop to zero consistent with the formation of the trivial ground state. But for the QSL and VBS phase, we could not reach temperatures low enough to reach the ground state:  They (locally) look the same at the temperatures we could reach~\footnote{Recall that the VBS consists of a localized subset of dimer configurations and can on small system sizes even have  an overlap with the RVB state  that is comparable to the one of the QSL with the RVB state.}.

The found entropy density of the plateau can be understood as follows. In the fully covered dimer limit, the degeneracy is exactly known~\cite{ZengElser1995,misguich_2002} to be $2^{N_{\rm hex} +1}$, with the number of hexagons $N_{\rm hex} = N/6$. This degeneracy is carried over to the low-energy manifold of a highly anisotropic $Z_2$ lattice gauge theory: If we think of a highly anisotropic toric code with periodic boundary conditions in which the magnetic excitations have a tiny energy compared to the electric ones, the spectrum has a manifold of low-energy states in which there are only magnetic excitations. The number of independent magnetic excitations is $2^{N_{\rm plaq}-1}$ with $N_{\rm plaq}$ the number of plaquettes. The global constraint stating that the product over all plaquette terms is always 1, takes one degree of freedom away. The total number of states in the low energy manifold is however $2^{N_{\rm plaq}+1}$ due to topology, exactly the same amount of states as in the fully covered dimer limit. 
The magnetic fluxes connect to the 32 resonating loop operations that can be defined for the dimer model~\cite{ZengElser1995,misguich_2002}, along which the dimers can be placed in two ways. Thus, $N_{\rm plaq} = N_{\rm hex}.$

Full diagonalization~\cite{Supplements} of the system with $L=2$ confirms that the entropy of the CSL is $S/N = \ln(2)(1/6 + 1/N)$ for any $L$. The smallness of the magnetic gap is caused by the absence of an explicit plaquette interaction and thus the fact that it has to be generated perturbatively by the Rabi term.
The lowest order operator allowing tunneling within the CSL  has amplitude ~\cite{verresen_prediction_2021}  $~\sim 3 \Omega^6/32 \delta^5$ in the large $\delta$ limit (there are also higher order processes). This scaling is also observed in full diagonalization~\cite{Supplements}.
If we naively plug in $\delta = 2$ such processes become relevant for $\beta \sim 300$ (in line with estimates from exact diagonalization on small system sizes~\cite{patil2023quantum, Supplements}), which is outside the regime we can simulate. 

\begin{figure}
    \centering
    \includegraphics[width=\columnwidth]{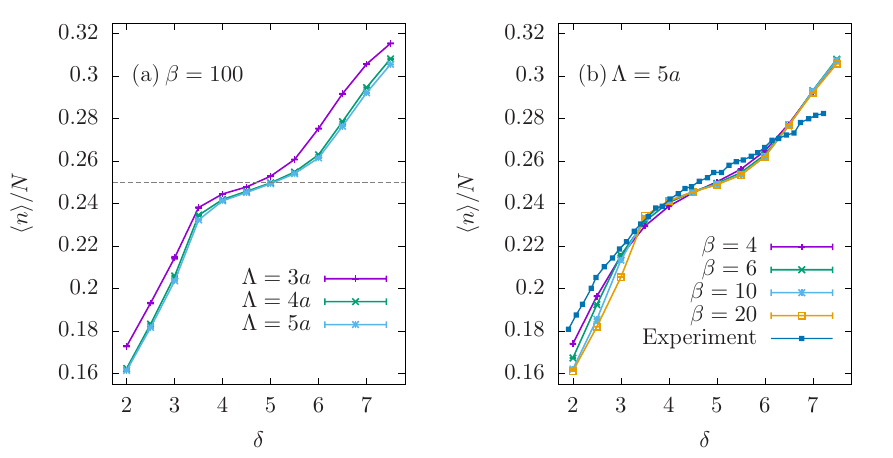}   
    \caption{ 
    Left: Density of Rydberg excitations for the system with VdW interactions truncated at (a) $A = 3a$, (b) $A = 4a$, and (c) $A = 5a$ at inverse temperature $\beta = 100$ for various values of the detuning $\delta$. Right: same as a function of inverse temperature at fixed truncation $A=5a$. Experimental data are taken from Ref.~\cite{semeghini_probing_2021}.} 
    \label{fig:density}
\end{figure}

{\it Van der Waals interactions -- } The observed phenomenology for the {\it PXP} model is qualitatively the same in the VdW model but there are noticeable quantitative changes.
As can be seen in Fig.~\ref{fig:entropy}(b), plateaus in the entropy density are seen over the relevant range $\delta=3.5$ to $\delta=4.5$ but with values that are less than $\log(2)/6$. A plateau with that value is seen for $\delta=5.5$ but this requires finetuning and is in a regime where DMRG finds a crystalline ground state. The plateaus are narrower and require lower temperature than for the {\it PXP} model. The fidelity susceptibility at low but finite temperature (Fig.~\ref{fig:fidelity}(b)) and the closed loop operator (Fig.~\ref{fig:Loop}(b)) confirm a transition from a trivial paramagnet to a spin liquid at $\delta \approx 3.5$ and a broad crossover towards a trivial phase at $\delta \approx 6.5$.  Thus, a CSL survives but the separation of electric and magnetic scales is less strong than in the {\it PXP} model (cf. Fig. S1 of \cite{Supplements}). The CSL shows perimeter scaling for the closed loop diagonal string parity operators, cf.~Fig.~\ref{fig:fidelity}(b) and Fig. S3 of~\cite{Supplements}. 

{\it Perspective on experiment --}
The observed two-scale behavior in the {\it PXP} and VdW models lends strong support to the spin-lake picture of Ref.~\cite{sahay_quantum_2023}: a dynamical ramp entering the QSL regime is sudden for the magnetic degrees of freedom but adiabatic for the electric ones, implying that the prepared magnetic ground state from the trivial phase survives initially in the SL regime and enables a perimeter scaling of the off-diagonal string parity operators.  A few electric excitations must however be present (with its degrees of freedom close to equilibrium) and the corresponding energy must be {\it higher} than the thermal energy of the CSL. 
Although it is unlikely that {\it full} equilibrium is reached for the electric degrees of freedom, we nevertheless find that the experimentally measured density of Rydberg excitations is compatible with the QMC results over the range $\delta=3.5$ to $\delta=4.5$ (see Fig.~\ref{fig:density}), putting an upper bound on the presumptive temperature,  $\beta \gtrsim 6$. 
The experimentally measured values for $\mathcal{Z}(l), l=2-4,$ are of similar magnitude as the QMC ones for temperatures in the range $\beta = 5-8$ (corresponding to $S/N \sim 0.15$), see~\cite{Supplements}.
The interpretation given in the experiment~\cite{semeghini_probing_2021}, namely a change from area to perimeter law with increasing $l$, is not compatible with a finite temperature equilibrium picture of the electric degrees of freedom: we find uniform area law scaling for $\beta < 10$.   For larger values of $\delta > 4.5$ we expect a rapid tendency towards a thermal paramagnet at temperatures slowly increasing with experimental duration possibly due to the increased role of  decoherence, cf. the situation in Refs.~\cite{Chen2023,Sbierski2024}.

{\it Conclusion and outlook -- }
Realizing quantum spin liquids through a Rydberg blockade mechanism implementing Gauss law is a topic of active research. The {\it PXP} model on the ruby lattice is a theoretical idealization of the VdW interactions and harbors such a QSL. Nevertheless, it also lays bare an inherent challenge for current experiments: in the absence of explicit plaquette terms, those must be generated perturbatively by the Rabi term, resulting in a magnetic energy scale that is orders of magnitude smaller than the electric one. At any energy (or temperature) scale in between, one can expect a renormalized classical spin liquid: the system is in the ground state with respect to the electric degrees of freedom, including quantum fluctuations caused by the Rabi term, but (almost) infinitely hot compared to the magnetic degrees of freedom (dimer resonances) resulting in an extensive entropy given by the degeneracy of the low-energy magnetic manifold. This picture is particularly strong in the {\it PXP} model.
The situation in current dynamical experiments is strikingly different: one attempts to freeze the magnetic ground state at the expense of a few costly electric excitations. Assuming equilibrium for the latter, the corresponding entropy density is $S/N \sim 0.12 - 0.15$.
As the conditions for a dynamically prepared QSL state and a ground-state QSL are difficult to reconcile and the ones for an equilibrium QSL especially challenging, studies to increase the lifetime and scalability of dynamically prepared QSL are promising avenues for further research.

We thank  I. Bloch, F. Grusdt, G. Giudici, L. Homeier, P. Patil, and N. Yao  for fruitful discussions.
This research was funded by the European Research Council (ERC) under the European Union’s Horizon 2020 research and innovation program (Grant Agreement no 771891) --  ERC Consolidator Grant QSIMCORR, and by the Deutsche Forschungsgemeinschaft (DFG, German Research Foundation) under Germany's Excellence Strategy -- EXC-2111 -- 390814868.

\bibliography{QSL}

\begin{thebibliography}{43}%
\makeatletter
\providecommand \@ifxundefined [1]{%
 \@ifx{#1\undefined}
}%
\providecommand \@ifnum [1]{%
 \ifnum #1\expandafter \@firstoftwo
 \else \expandafter \@secondoftwo
 \fi
}%
\providecommand \@ifx [1]{%
 \ifx #1\expandafter \@firstoftwo
 \else \expandafter \@secondoftwo
 \fi
}%
\providecommand \natexlab [1]{#1}%
\providecommand \enquote  [1]{``#1''}%
\providecommand \bibnamefont  [1]{#1}%
\providecommand \bibfnamefont [1]{#1}%
\providecommand \citenamefont [1]{#1}%
\providecommand \href@noop [0]{\@secondoftwo}%
\providecommand \href [0]{\begingroup \@sanitize@url \@href}%
\providecommand \@href[1]{\@@startlink{#1}\@@href}%
\providecommand \@@href[1]{\endgroup#1\@@endlink}%
\providecommand \@sanitize@url [0]{\catcode `\\12\catcode `\$12\catcode
  `\&12\catcode `\#12\catcode `\^12\catcode `\_12\catcode `\%12\relax}%
\providecommand \@@startlink[1]{}%
\providecommand \@@endlink[0]{}%
\providecommand \url  [0]{\begingroup\@sanitize@url \@url }%
\providecommand \@url [1]{\endgroup\@href {#1}{\urlprefix }}%
\providecommand \urlprefix  [0]{URL }%
\providecommand \Eprint [0]{\href }%
\providecommand \doibase [0]{https://doi.org/}%
\providecommand \selectlanguage [0]{\@gobble}%
\providecommand \bibinfo  [0]{\@secondoftwo}%
\providecommand \bibfield  [0]{\@secondoftwo}%
\providecommand \translation [1]{[#1]}%
\providecommand \BibitemOpen [0]{}%
\providecommand \bibitemStop [0]{}%
\providecommand \bibitemNoStop [0]{.\EOS\space}%
\providecommand \EOS [0]{\spacefactor3000\relax}%
\providecommand \BibitemShut  [1]{\csname bibitem#1\endcsname}%
\let\auto@bib@innerbib\@empty
\bibitem [{\citenamefont {Semeghini}\ \emph {et~al.}(2021)\citenamefont
  {Semeghini}, \citenamefont {Levine}, \citenamefont {Keesling}, \citenamefont
  {Ebadi}, \citenamefont {Wang}, \citenamefont {Bluvstein}, \citenamefont
  {Verresen}, \citenamefont {Pichler}, \citenamefont {Kalinowski},
  \citenamefont {Samajdar}, \citenamefont {Omran}, \citenamefont {Sachdev},
  \citenamefont {Vishwanath}, \citenamefont {Greiner}, \citenamefont
  {Vuletic},\ and\ \citenamefont {Lukin}}]{semeghini_probing_2021}%
  \BibitemOpen
  \bibfield  {author} {\bibinfo {author} {\bibfnamefont {G.}~\bibnamefont
  {Semeghini}}, \bibinfo {author} {\bibfnamefont {H.}~\bibnamefont {Levine}},
  \bibinfo {author} {\bibfnamefont {A.}~\bibnamefont {Keesling}}, \bibinfo
  {author} {\bibfnamefont {S.}~\bibnamefont {Ebadi}}, \bibinfo {author}
  {\bibfnamefont {T.~T.}\ \bibnamefont {Wang}}, \bibinfo {author}
  {\bibfnamefont {D.}~\bibnamefont {Bluvstein}}, \bibinfo {author}
  {\bibfnamefont {R.}~\bibnamefont {Verresen}}, \bibinfo {author}
  {\bibfnamefont {H.}~\bibnamefont {Pichler}}, \bibinfo {author} {\bibfnamefont
  {M.}~\bibnamefont {Kalinowski}}, \bibinfo {author} {\bibfnamefont
  {R.}~\bibnamefont {Samajdar}}, \bibinfo {author} {\bibfnamefont
  {A.}~\bibnamefont {Omran}}, \bibinfo {author} {\bibfnamefont
  {S.}~\bibnamefont {Sachdev}}, \bibinfo {author} {\bibfnamefont
  {A.}~\bibnamefont {Vishwanath}}, \bibinfo {author} {\bibfnamefont
  {M.}~\bibnamefont {Greiner}}, \bibinfo {author} {\bibfnamefont
  {V.}~\bibnamefont {Vuletic}},\ and\ \bibinfo {author} {\bibfnamefont {M.~D.}\
  \bibnamefont {Lukin}},\ }\bibfield  {title} {\bibinfo {title} {Probing
  {Topological} {Spin} {Liquids} on a {Programmable} {Quantum} {Simulator}},\
  }\href {https://doi.org/10.1126/science.abi8794} {\bibfield  {journal}
  {\bibinfo  {journal} {Science}\ }\textbf {\bibinfo {volume} {374}},\ \bibinfo
  {pages} {1242} (\bibinfo {year} {2021})},\ \bibinfo {note} {arXiv:2104.04119
  [cond-mat, physics:physics, physics:quant-ph]}\BibitemShut {NoStop}%
\bibitem [{\citenamefont {Savary}\ and\ \citenamefont
  {Balents}(2017)}]{savary_quantum_2017}%
  \BibitemOpen
  \bibfield  {author} {\bibinfo {author} {\bibfnamefont {L.}~\bibnamefont
  {Savary}}\ and\ \bibinfo {author} {\bibfnamefont {L.}~\bibnamefont
  {Balents}},\ }\bibfield  {title} {\bibinfo {title} {Quantum spin liquids: a
  review},\ }\href {https://doi.org/10.1088/0034-4885/80/1/016502} {\bibfield
  {journal} {\bibinfo  {journal} {Rep. Prog. Phys.}\ }\textbf {\bibinfo
  {volume} {80}},\ \bibinfo {pages} {016502} (\bibinfo {year}
  {2017})}\BibitemShut {NoStop}%
\bibitem [{\citenamefont {Kitaev}(2003)}]{kitaev_2003}%
  \BibitemOpen
  \bibfield  {author} {\bibinfo {author} {\bibfnamefont {A.}~\bibnamefont
  {Kitaev}},\ }\bibfield  {title} {\bibinfo {title} {Fault-tolerant quantum
  computation by anyons},\ }\href
  {https://doi.org/https://doi.org/10.1016/S0003-4916(02)00018-0} {\bibfield
  {journal} {\bibinfo  {journal} {Annals of Physics}\ }\textbf {\bibinfo
  {volume} {303}},\ \bibinfo {pages} {2} (\bibinfo {year} {2003})}\BibitemShut
  {NoStop}%
\bibitem [{\citenamefont {Wen}(2017)}]{wen_colloquium_2017}%
  \BibitemOpen
  \bibfield  {author} {\bibinfo {author} {\bibfnamefont {X.-G.}\ \bibnamefont
  {Wen}},\ }\bibfield  {title} {\bibinfo {title} {Colloquium: Zoo of
  quantum-topological phases of matter},\ }\href
  {https://doi.org/10.1103/RevModPhys.89.041004} {\bibfield  {journal}
  {\bibinfo  {journal} {Rev. Mod. Phys.}\ }\textbf {\bibinfo {volume} {89}},\
  \bibinfo {pages} {041004} (\bibinfo {year} {2017})}\BibitemShut {NoStop}%
\bibitem [{\citenamefont {Wilczek}(1982)}]{wilczek_fractional_1982}%
  \BibitemOpen
  \bibfield  {author} {\bibinfo {author} {\bibfnamefont {F.}~\bibnamefont
  {Wilczek}},\ }\bibfield  {title} {\bibinfo {title} {Quantum mechanics of
  fractional-spin particles},\ }\href
  {https://doi.org/10.1103/PhysRevLett.49.957} {\bibfield  {journal} {\bibinfo
  {journal} {Phys. Rev. Lett.}\ }\textbf {\bibinfo {volume} {49}},\ \bibinfo
  {pages} {957} (\bibinfo {year} {1982})}\BibitemShut {NoStop}%
\bibitem [{\citenamefont {Nayak}\ \emph {et~al.}(2008)\citenamefont {Nayak},
  \citenamefont {Simon}, \citenamefont {Stern}, \citenamefont {Freedman},\ and\
  \citenamefont {Das~Sarma}}]{nayak_rmp_2008}%
  \BibitemOpen
  \bibfield  {author} {\bibinfo {author} {\bibfnamefont {C.}~\bibnamefont
  {Nayak}}, \bibinfo {author} {\bibfnamefont {S.~H.}\ \bibnamefont {Simon}},
  \bibinfo {author} {\bibfnamefont {A.}~\bibnamefont {Stern}}, \bibinfo
  {author} {\bibfnamefont {M.}~\bibnamefont {Freedman}},\ and\ \bibinfo
  {author} {\bibfnamefont {S.}~\bibnamefont {Das~Sarma}},\ }\bibfield  {title}
  {\bibinfo {title} {Non-abelian anyons and topological quantum computation},\
  }\href {https://doi.org/10.1103/RevModPhys.80.1083} {\bibfield  {journal}
  {\bibinfo  {journal} {Rev. Mod. Phys.}\ }\textbf {\bibinfo {volume} {80}},\
  \bibinfo {pages} {1083} (\bibinfo {year} {2008})}\BibitemShut {NoStop}%
\bibitem [{\citenamefont {Terhal}(2015)}]{terhal_rmp_2015}%
  \BibitemOpen
  \bibfield  {author} {\bibinfo {author} {\bibfnamefont {B.~M.}\ \bibnamefont
  {Terhal}},\ }\bibfield  {title} {\bibinfo {title} {Quantum error correction
  for quantum memories},\ }\href {https://doi.org/10.1103/RevModPhys.87.307}
  {\bibfield  {journal} {\bibinfo  {journal} {Rev. Mod. Phys.}\ }\textbf
  {\bibinfo {volume} {87}},\ \bibinfo {pages} {307} (\bibinfo {year}
  {2015})}\BibitemShut {NoStop}%
\bibitem [{\citenamefont {Satzinger}\ \emph {et~al.}(2021)\citenamefont
  {Satzinger}, \citenamefont {Liu}, \citenamefont {Smith}, \citenamefont
  {Knapp}, \citenamefont {Newman}, \citenamefont {Jones}, \citenamefont {Chen},
  \citenamefont {Quintana}, \citenamefont {Mi}, \citenamefont {Dunsworth},
  \citenamefont {Gidney}, \citenamefont {Aleiner}, \citenamefont {Arute},
  \citenamefont {Arya}, \citenamefont {Atalaya}, \citenamefont {Babbush},
  \citenamefont {Bardin}, \citenamefont {Barends}, \citenamefont {Basso},
  \citenamefont {Bengtsson}, \citenamefont {Bilmes}, \citenamefont {Broughton},
  \citenamefont {Buckley}, \citenamefont {Buell}, \citenamefont {Burkett},
  \citenamefont {Bushnell}, \citenamefont {Chiaro}, \citenamefont {Collins},
  \citenamefont {Courtney}, \citenamefont {Demura}, \citenamefont {Derk},
  \citenamefont {Eppens}, \citenamefont {Erickson}, \citenamefont {Faoro},
  \citenamefont {Farhi}, \citenamefont {Fowler}, \citenamefont {Foxen},
  \citenamefont {Giustina}, \citenamefont {Greene}, \citenamefont {Gross},
  \citenamefont {Harrigan}, \citenamefont {Harrington}, \citenamefont {Hilton},
  \citenamefont {Hong}, \citenamefont {Huang}, \citenamefont {Huggins},
  \citenamefont {Ioffe}, \citenamefont {Isakov}, \citenamefont {Jeffrey},
  \citenamefont {Jiang}, \citenamefont {Kafri}, \citenamefont {Kechedzhi},
  \citenamefont {Khattar}, \citenamefont {Kim}, \citenamefont {Klimov},
  \citenamefont {Korotkov}, \citenamefont {Kostritsa}, \citenamefont
  {Landhuis}, \citenamefont {Laptev}, \citenamefont {Locharla}, \citenamefont
  {Lucero}, \citenamefont {Martin}, \citenamefont {McClean}, \citenamefont
  {McEwen}, \citenamefont {Miao}, \citenamefont {Mohseni}, \citenamefont
  {Montazeri}, \citenamefont {Mruczkiewicz}, \citenamefont {Mutus},
  \citenamefont {Naaman}, \citenamefont {Neeley}, \citenamefont {Neill},
  \citenamefont {Niu}, \citenamefont {O’Brien}, \citenamefont {Opremcak},
  \citenamefont {Pató}, \citenamefont {Petukhov}, \citenamefont {Rubin},
  \citenamefont {Sank}, \citenamefont {Shvarts}, \citenamefont {Strain},
  \citenamefont {Szalay}, \citenamefont {Villalonga}, \citenamefont {White},
  \citenamefont {Yao}, \citenamefont {Yeh}, \citenamefont {Yoo}, \citenamefont
  {Zalcman}, \citenamefont {Neven}, \citenamefont {Boixo}, \citenamefont
  {Megrant}, \citenamefont {Chen}, \citenamefont {Kelly}, \citenamefont
  {Smelyanskiy}, \citenamefont {Kitaev}, \citenamefont {Knap}, \citenamefont
  {Pollmann},\ and\ \citenamefont {Roushan}}]{satzinger_toric_2021}%
  \BibitemOpen
  \bibfield  {author} {\bibinfo {author} {\bibfnamefont {K.~J.}\ \bibnamefont
  {Satzinger}}, \bibinfo {author} {\bibfnamefont {Y.-J.}\ \bibnamefont {Liu}},
  \bibinfo {author} {\bibfnamefont {A.}~\bibnamefont {Smith}}, \bibinfo
  {author} {\bibfnamefont {C.}~\bibnamefont {Knapp}}, \bibinfo {author}
  {\bibfnamefont {M.}~\bibnamefont {Newman}}, \bibinfo {author} {\bibfnamefont
  {C.}~\bibnamefont {Jones}}, \bibinfo {author} {\bibfnamefont
  {Z.}~\bibnamefont {Chen}}, \bibinfo {author} {\bibfnamefont {C.}~\bibnamefont
  {Quintana}}, \bibinfo {author} {\bibfnamefont {X.}~\bibnamefont {Mi}},
  \bibinfo {author} {\bibfnamefont {A.}~\bibnamefont {Dunsworth}}, \bibinfo
  {author} {\bibfnamefont {C.}~\bibnamefont {Gidney}}, \bibinfo {author}
  {\bibfnamefont {I.}~\bibnamefont {Aleiner}}, \bibinfo {author} {\bibfnamefont
  {F.}~\bibnamefont {Arute}}, \bibinfo {author} {\bibfnamefont
  {K.}~\bibnamefont {Arya}}, \bibinfo {author} {\bibfnamefont {J.}~\bibnamefont
  {Atalaya}}, \bibinfo {author} {\bibfnamefont {R.}~\bibnamefont {Babbush}},
  \bibinfo {author} {\bibfnamefont {J.~C.}\ \bibnamefont {Bardin}}, \bibinfo
  {author} {\bibfnamefont {R.}~\bibnamefont {Barends}}, \bibinfo {author}
  {\bibfnamefont {J.}~\bibnamefont {Basso}}, \bibinfo {author} {\bibfnamefont
  {A.}~\bibnamefont {Bengtsson}}, \bibinfo {author} {\bibfnamefont
  {A.}~\bibnamefont {Bilmes}}, \bibinfo {author} {\bibfnamefont
  {M.}~\bibnamefont {Broughton}}, \bibinfo {author} {\bibfnamefont {B.~B.}\
  \bibnamefont {Buckley}}, \bibinfo {author} {\bibfnamefont {D.~A.}\
  \bibnamefont {Buell}}, \bibinfo {author} {\bibfnamefont {B.}~\bibnamefont
  {Burkett}}, \bibinfo {author} {\bibfnamefont {N.}~\bibnamefont {Bushnell}},
  \bibinfo {author} {\bibfnamefont {B.}~\bibnamefont {Chiaro}}, \bibinfo
  {author} {\bibfnamefont {R.}~\bibnamefont {Collins}}, \bibinfo {author}
  {\bibfnamefont {W.}~\bibnamefont {Courtney}}, \bibinfo {author}
  {\bibfnamefont {S.}~\bibnamefont {Demura}}, \bibinfo {author} {\bibfnamefont
  {A.~R.}\ \bibnamefont {Derk}}, \bibinfo {author} {\bibfnamefont
  {D.}~\bibnamefont {Eppens}}, \bibinfo {author} {\bibfnamefont
  {C.}~\bibnamefont {Erickson}}, \bibinfo {author} {\bibfnamefont
  {L.}~\bibnamefont {Faoro}}, \bibinfo {author} {\bibfnamefont
  {E.}~\bibnamefont {Farhi}}, \bibinfo {author} {\bibfnamefont {A.~G.}\
  \bibnamefont {Fowler}}, \bibinfo {author} {\bibfnamefont {B.}~\bibnamefont
  {Foxen}}, \bibinfo {author} {\bibfnamefont {M.}~\bibnamefont {Giustina}},
  \bibinfo {author} {\bibfnamefont {A.}~\bibnamefont {Greene}}, \bibinfo
  {author} {\bibfnamefont {J.~A.}\ \bibnamefont {Gross}}, \bibinfo {author}
  {\bibfnamefont {M.~P.}\ \bibnamefont {Harrigan}}, \bibinfo {author}
  {\bibfnamefont {S.~D.}\ \bibnamefont {Harrington}}, \bibinfo {author}
  {\bibfnamefont {J.}~\bibnamefont {Hilton}}, \bibinfo {author} {\bibfnamefont
  {S.}~\bibnamefont {Hong}}, \bibinfo {author} {\bibfnamefont {T.}~\bibnamefont
  {Huang}}, \bibinfo {author} {\bibfnamefont {W.~J.}\ \bibnamefont {Huggins}},
  \bibinfo {author} {\bibfnamefont {L.~B.}\ \bibnamefont {Ioffe}}, \bibinfo
  {author} {\bibfnamefont {S.~V.}\ \bibnamefont {Isakov}}, \bibinfo {author}
  {\bibfnamefont {E.}~\bibnamefont {Jeffrey}}, \bibinfo {author} {\bibfnamefont
  {Z.}~\bibnamefont {Jiang}}, \bibinfo {author} {\bibfnamefont
  {D.}~\bibnamefont {Kafri}}, \bibinfo {author} {\bibfnamefont
  {K.}~\bibnamefont {Kechedzhi}}, \bibinfo {author} {\bibfnamefont
  {T.}~\bibnamefont {Khattar}}, \bibinfo {author} {\bibfnamefont
  {S.}~\bibnamefont {Kim}}, \bibinfo {author} {\bibfnamefont {P.~V.}\
  \bibnamefont {Klimov}}, \bibinfo {author} {\bibfnamefont {A.~N.}\
  \bibnamefont {Korotkov}}, \bibinfo {author} {\bibfnamefont {F.}~\bibnamefont
  {Kostritsa}}, \bibinfo {author} {\bibfnamefont {D.}~\bibnamefont {Landhuis}},
  \bibinfo {author} {\bibfnamefont {P.}~\bibnamefont {Laptev}}, \bibinfo
  {author} {\bibfnamefont {A.}~\bibnamefont {Locharla}}, \bibinfo {author}
  {\bibfnamefont {E.}~\bibnamefont {Lucero}}, \bibinfo {author} {\bibfnamefont
  {O.}~\bibnamefont {Martin}}, \bibinfo {author} {\bibfnamefont {J.~R.}\
  \bibnamefont {McClean}}, \bibinfo {author} {\bibfnamefont {M.}~\bibnamefont
  {McEwen}}, \bibinfo {author} {\bibfnamefont {K.~C.}\ \bibnamefont {Miao}},
  \bibinfo {author} {\bibfnamefont {M.}~\bibnamefont {Mohseni}}, \bibinfo
  {author} {\bibfnamefont {S.}~\bibnamefont {Montazeri}}, \bibinfo {author}
  {\bibfnamefont {W.}~\bibnamefont {Mruczkiewicz}}, \bibinfo {author}
  {\bibfnamefont {J.}~\bibnamefont {Mutus}}, \bibinfo {author} {\bibfnamefont
  {O.}~\bibnamefont {Naaman}}, \bibinfo {author} {\bibfnamefont
  {M.}~\bibnamefont {Neeley}}, \bibinfo {author} {\bibfnamefont
  {C.}~\bibnamefont {Neill}}, \bibinfo {author} {\bibfnamefont {M.~Y.}\
  \bibnamefont {Niu}}, \bibinfo {author} {\bibfnamefont {T.~E.}\ \bibnamefont
  {O’Brien}}, \bibinfo {author} {\bibfnamefont {A.}~\bibnamefont {Opremcak}},
  \bibinfo {author} {\bibfnamefont {B.}~\bibnamefont {Pató}}, \bibinfo
  {author} {\bibfnamefont {A.}~\bibnamefont {Petukhov}}, \bibinfo {author}
  {\bibfnamefont {N.~C.}\ \bibnamefont {Rubin}}, \bibinfo {author}
  {\bibfnamefont {D.}~\bibnamefont {Sank}}, \bibinfo {author} {\bibfnamefont
  {V.}~\bibnamefont {Shvarts}}, \bibinfo {author} {\bibfnamefont
  {D.}~\bibnamefont {Strain}}, \bibinfo {author} {\bibfnamefont
  {M.}~\bibnamefont {Szalay}}, \bibinfo {author} {\bibfnamefont
  {B.}~\bibnamefont {Villalonga}}, \bibinfo {author} {\bibfnamefont {T.~C.}\
  \bibnamefont {White}}, \bibinfo {author} {\bibfnamefont {Z.}~\bibnamefont
  {Yao}}, \bibinfo {author} {\bibfnamefont {P.}~\bibnamefont {Yeh}}, \bibinfo
  {author} {\bibfnamefont {J.}~\bibnamefont {Yoo}}, \bibinfo {author}
  {\bibfnamefont {A.}~\bibnamefont {Zalcman}}, \bibinfo {author} {\bibfnamefont
  {H.}~\bibnamefont {Neven}}, \bibinfo {author} {\bibfnamefont
  {S.}~\bibnamefont {Boixo}}, \bibinfo {author} {\bibfnamefont
  {A.}~\bibnamefont {Megrant}}, \bibinfo {author} {\bibfnamefont
  {Y.}~\bibnamefont {Chen}}, \bibinfo {author} {\bibfnamefont {J.}~\bibnamefont
  {Kelly}}, \bibinfo {author} {\bibfnamefont {V.}~\bibnamefont {Smelyanskiy}},
  \bibinfo {author} {\bibfnamefont {A.}~\bibnamefont {Kitaev}}, \bibinfo
  {author} {\bibfnamefont {M.}~\bibnamefont {Knap}}, \bibinfo {author}
  {\bibfnamefont {F.}~\bibnamefont {Pollmann}},\ and\ \bibinfo {author}
  {\bibfnamefont {P.}~\bibnamefont {Roushan}},\ }\bibfield  {title} {\bibinfo
  {title} {Realizing topologically ordered states on a quantum processor},\
  }\href {https://doi.org/10.1126/science.abi8378} {\bibfield  {journal}
  {\bibinfo  {journal} {Science}\ }\textbf {\bibinfo {volume} {374}},\ \bibinfo
  {pages} {1237} (\bibinfo {year} {2021})},\ \Eprint
  {https://arxiv.org/abs/https://www.science.org/doi/pdf/10.1126/science.abi8378}
  {https://www.science.org/doi/pdf/10.1126/science.abi8378} \BibitemShut
  {NoStop}%
\bibitem [{\citenamefont {Wegner}(1971)}]{wegner_1971}%
  \BibitemOpen
  \bibfield  {author} {\bibinfo {author} {\bibfnamefont {F.~J.}\ \bibnamefont
  {Wegner}},\ }\bibfield  {title} {\bibinfo {title} {{Duality in Generalized
  Ising Models and Phase Transitions without Local Order Parameters}},\ }\href
  {https://doi.org/10.1063/1.1665530} {\bibfield  {journal} {\bibinfo
  {journal} {Journal of Mathematical Physics}\ }\textbf {\bibinfo {volume}
  {12}},\ \bibinfo {pages} {2259} (\bibinfo {year} {1971})},\ \Eprint
  {https://arxiv.org/abs/https://pubs.aip.org/aip/jmp/article-pdf/12/10/2259/19106483/2259\_1\_online.pdf}
  {https://pubs.aip.org/aip/jmp/article-pdf/12/10/2259/19106483/2259\_1\_online.pdf}
  \BibitemShut {NoStop}%
\bibitem [{\citenamefont {Fradkin}\ and\ \citenamefont
  {Shenker}(1979)}]{fradkin_phase_1979}%
  \BibitemOpen
  \bibfield  {author} {\bibinfo {author} {\bibfnamefont {E.}~\bibnamefont
  {Fradkin}}\ and\ \bibinfo {author} {\bibfnamefont {S.~H.}\ \bibnamefont
  {Shenker}},\ }\bibfield  {title} {\bibinfo {title} {Phase diagrams of lattice
  gauge theories with {Higgs} fields},\ }\href
  {https://doi.org/10.1103/PhysRevD.19.3682} {\bibfield  {journal} {\bibinfo
  {journal} {Phys. Rev. D}\ }\textbf {\bibinfo {volume} {19}},\ \bibinfo
  {pages} {3682} (\bibinfo {year} {1979})}\BibitemShut {NoStop}%
\bibitem [{\citenamefont {Kitaev}(2006)}]{kitaev_honeycomb_2006}%
  \BibitemOpen
  \bibfield  {author} {\bibinfo {author} {\bibfnamefont {A.}~\bibnamefont
  {Kitaev}},\ }\bibfield  {title} {\bibinfo {title} {Anyons in an exactly
  solved model and beyond},\ }\href
  {https://doi.org/https://doi.org/10.1016/j.aop.2005.10.005} {\bibfield
  {journal} {\bibinfo  {journal} {Annals of Physics}\ }\textbf {\bibinfo
  {volume} {321}},\ \bibinfo {pages} {2} (\bibinfo {year} {2006})}\BibitemShut
  {NoStop}%
\bibitem [{\citenamefont {Rokhsar}\ and\ \citenamefont
  {Kivelson}(1988)}]{rokhsar_superconductivity_1988}%
  \BibitemOpen
  \bibfield  {author} {\bibinfo {author} {\bibfnamefont {D.~S.}\ \bibnamefont
  {Rokhsar}}\ and\ \bibinfo {author} {\bibfnamefont {S.~A.}\ \bibnamefont
  {Kivelson}},\ }\bibfield  {title} {\bibinfo {title} {Superconductivity and
  the {Quantum} {Hard}-{Core} {Dimer} {Gas}},\ }\href
  {https://doi.org/10.1103/PhysRevLett.61.2376} {\bibfield  {journal} {\bibinfo
   {journal} {Phys. Rev. Lett.}\ }\textbf {\bibinfo {volume} {61}},\ \bibinfo
  {pages} {2376} (\bibinfo {year} {1988})}\BibitemShut {NoStop}%
\bibitem [{\citenamefont {Sachdev}(1992)}]{sachdev_1992}%
  \BibitemOpen
  \bibfield  {author} {\bibinfo {author} {\bibfnamefont {S.}~\bibnamefont
  {Sachdev}},\ }\bibfield  {title} {\bibinfo {title} {Kagom{\'e}- and
  triangular-lattice heisenberg antiferromagnets: Ordering from quantum
  fluctuations and quantum-disordered ground states with unconfined bosonic
  spinons},\ }\href {https://doi.org/10.1103/PhysRevB.45.12377} {\bibfield
  {journal} {\bibinfo  {journal} {Phys. Rev. B}\ }\textbf {\bibinfo {volume}
  {45}},\ \bibinfo {pages} {12377} (\bibinfo {year} {1992})}\BibitemShut
  {NoStop}%
\bibitem [{\citenamefont {Moessner}\ and\ \citenamefont
  {Raman}(2008)}]{moessner_quantum_2008}%
  \BibitemOpen
  \bibfield  {author} {\bibinfo {author} {\bibfnamefont {R.}~\bibnamefont
  {Moessner}}\ and\ \bibinfo {author} {\bibfnamefont {K.~S.}\ \bibnamefont
  {Raman}},\ }\href {http://arxiv.org/abs/0809.3051} {\bibinfo {title} {Quantum
  dimer models}} (\bibinfo {year} {2008}),\ \bibinfo {note} {arXiv:0809.3051
  [cond-mat]}\BibitemShut {NoStop}%
\bibitem [{\citenamefont {Moessner}\ \emph {et~al.}(2001)\citenamefont
  {Moessner}, \citenamefont {Sondhi},\ and\ \citenamefont
  {Fradkin}}]{moessner_2001}%
  \BibitemOpen
  \bibfield  {author} {\bibinfo {author} {\bibfnamefont {R.}~\bibnamefont
  {Moessner}}, \bibinfo {author} {\bibfnamefont {S.~L.}\ \bibnamefont
  {Sondhi}},\ and\ \bibinfo {author} {\bibfnamefont {E.}~\bibnamefont
  {Fradkin}},\ }\bibfield  {title} {\bibinfo {title} {Short-ranged resonating
  valence bond physics, quantum dimer models, and {I}sing gauge theories},\
  }\href {https://doi.org/10.1103/PhysRevB.65.024504} {\bibfield  {journal}
  {\bibinfo  {journal} {Phys. Rev. B}\ }\textbf {\bibinfo {volume} {65}},\
  \bibinfo {pages} {024504} (\bibinfo {year} {2001})}\BibitemShut {NoStop}%
\bibitem [{\citenamefont {Misguich}\ \emph {et~al.}(2002)\citenamefont
  {Misguich}, \citenamefont {Serban},\ and\ \citenamefont
  {Pasquier}}]{misguich_2002}%
  \BibitemOpen
  \bibfield  {author} {\bibinfo {author} {\bibfnamefont {G.}~\bibnamefont
  {Misguich}}, \bibinfo {author} {\bibfnamefont {D.}~\bibnamefont {Serban}},\
  and\ \bibinfo {author} {\bibfnamefont {V.}~\bibnamefont {Pasquier}},\
  }\bibfield  {title} {\bibinfo {title} {Quantum dimer model on the kagome
  lattice: Solvable dimer-liquid and {I}sing gauge theory},\ }\href
  {https://doi.org/10.1103/PhysRevLett.89.137202} {\bibfield  {journal}
  {\bibinfo  {journal} {Phys. Rev. Lett.}\ }\textbf {\bibinfo {volume} {89}},\
  \bibinfo {pages} {137202} (\bibinfo {year} {2002})}\BibitemShut {NoStop}%
\bibitem [{\citenamefont {Baskaran}\ \emph {et~al.}(2008)\citenamefont
  {Baskaran}, \citenamefont {Sen},\ and\ \citenamefont
  {Shankar}}]{Baskaran_2008}%
  \BibitemOpen
  \bibfield  {author} {\bibinfo {author} {\bibfnamefont {G.}~\bibnamefont
  {Baskaran}}, \bibinfo {author} {\bibfnamefont {D.}~\bibnamefont {Sen}},\ and\
  \bibinfo {author} {\bibfnamefont {R.}~\bibnamefont {Shankar}},\ }\bibfield
  {title} {\bibinfo {title} {Spin-$s$ kitaev model: Classical ground states,
  order from disorder, and exact correlation functions},\ }\href
  {https://doi.org/10.1103/PhysRevB.78.115116} {\bibfield  {journal} {\bibinfo
  {journal} {Phys. Rev. B}\ }\textbf {\bibinfo {volume} {78}},\ \bibinfo
  {pages} {115116} (\bibinfo {year} {2008})}\BibitemShut {NoStop}%
\bibitem [{\citenamefont {Rousochatzakis}\ \emph {et~al.}(2018)\citenamefont
  {Rousochatzakis}, \citenamefont {Sizyuk},\ and\ \citenamefont
  {Perkins}}]{Rousochatzakis2018}%
  \BibitemOpen
  \bibfield  {author} {\bibinfo {author} {\bibfnamefont {I.}~\bibnamefont
  {Rousochatzakis}}, \bibinfo {author} {\bibfnamefont {Y.}~\bibnamefont
  {Sizyuk}},\ and\ \bibinfo {author} {\bibfnamefont {N.~B.}\ \bibnamefont
  {Perkins}},\ }\bibfield  {title} {\bibinfo {title} {Quantum spin liquid in
  the semiclassical regime},\ }\href
  {https://doi.org/10.1038/s41467-018-03934-1} {\bibfield  {journal} {\bibinfo
  {journal} {Nature Communications}\ }\textbf {\bibinfo {volume} {9}},\
  \bibinfo {pages} {1575} (\bibinfo {year} {2018})}\BibitemShut {NoStop}%
\bibitem [{\citenamefont {Chen}\ \emph {et~al.}(2022)\citenamefont {Chen},
  \citenamefont {Genzor}, \citenamefont {Kim},\ and\ \citenamefont
  {Kao}}]{chen_spinone_kitaev_2022}%
  \BibitemOpen
  \bibfield  {author} {\bibinfo {author} {\bibfnamefont {Y.-H.}\ \bibnamefont
  {Chen}}, \bibinfo {author} {\bibfnamefont {J.}~\bibnamefont {Genzor}},
  \bibinfo {author} {\bibfnamefont {Y.~B.}\ \bibnamefont {Kim}},\ and\ \bibinfo
  {author} {\bibfnamefont {Y.-J.}\ \bibnamefont {Kao}},\ }\bibfield  {title}
  {\bibinfo {title} {Excitation spectrum of spin-1 kitaev spin liquids},\
  }\href {https://doi.org/10.1103/PhysRevB.105.L060403} {\bibfield  {journal}
  {\bibinfo  {journal} {Phys. Rev. B}\ }\textbf {\bibinfo {volume} {105}},\
  \bibinfo {pages} {L060403} (\bibinfo {year} {2022})}\BibitemShut {NoStop}%
\bibitem [{\citenamefont {Verresen}\ \emph {et~al.}(2021)\citenamefont
  {Verresen}, \citenamefont {Lukin},\ and\ \citenamefont
  {Vishwanath}}]{verresen_prediction_2021}%
  \BibitemOpen
  \bibfield  {author} {\bibinfo {author} {\bibfnamefont {R.}~\bibnamefont
  {Verresen}}, \bibinfo {author} {\bibfnamefont {M.~D.}\ \bibnamefont
  {Lukin}},\ and\ \bibinfo {author} {\bibfnamefont {A.}~\bibnamefont
  {Vishwanath}},\ }\bibfield  {title} {\bibinfo {title} {Prediction of {Toric}
  {Code} {Topological} {Order} from {Rydberg} {Blockade}},\ }\href
  {https://doi.org/10.1103/PhysRevX.11.031005} {\bibfield  {journal} {\bibinfo
  {journal} {Phys. Rev. X}\ }\textbf {\bibinfo {volume} {11}},\ \bibinfo
  {pages} {031005} (\bibinfo {year} {2021})}\BibitemShut {NoStop}%
\bibitem [{\citenamefont {Bernien}\ \emph {et~al.}(2017)\citenamefont
  {Bernien}, \citenamefont {Schwartz}, \citenamefont {Keesling}, \citenamefont
  {Levine}, \citenamefont {Omran}, \citenamefont {Pichler}, \citenamefont
  {Choi}, \citenamefont {Zibrov}, \citenamefont {Endres}, \citenamefont
  {Greiner}, \citenamefont {Vuleti{\'{c}}},\ and\ \citenamefont
  {Lukin}}]{bernien_probing_2017}%
  \BibitemOpen
  \bibfield  {author} {\bibinfo {author} {\bibfnamefont {H.}~\bibnamefont
  {Bernien}}, \bibinfo {author} {\bibfnamefont {S.}~\bibnamefont {Schwartz}},
  \bibinfo {author} {\bibfnamefont {A.}~\bibnamefont {Keesling}}, \bibinfo
  {author} {\bibfnamefont {H.}~\bibnamefont {Levine}}, \bibinfo {author}
  {\bibfnamefont {A.}~\bibnamefont {Omran}}, \bibinfo {author} {\bibfnamefont
  {H.}~\bibnamefont {Pichler}}, \bibinfo {author} {\bibfnamefont
  {S.}~\bibnamefont {Choi}}, \bibinfo {author} {\bibfnamefont {A.~S.}\
  \bibnamefont {Zibrov}}, \bibinfo {author} {\bibfnamefont {M.}~\bibnamefont
  {Endres}}, \bibinfo {author} {\bibfnamefont {M.}~\bibnamefont {Greiner}},
  \bibinfo {author} {\bibfnamefont {V.}~\bibnamefont {Vuleti{\'{c}}}},\ and\
  \bibinfo {author} {\bibfnamefont {M.~D.}\ \bibnamefont {Lukin}},\ }\bibfield
  {title} {\bibinfo {title} {Probing many-body dynamics on a 51-atom quantum
  simulator},\ }\href {https://doi.org/10.1038/nature24622} {\bibfield
  {journal} {\bibinfo  {journal} {Nature}\ }\textbf {\bibinfo {volume} {551}},\
  \bibinfo {pages} {579} (\bibinfo {year} {2017})}\BibitemShut {NoStop}%
\bibitem [{\citenamefont {Browaeys}\ and\ \citenamefont
  {Lahaye}(2020)}]{browaeys_many-body_2020}%
  \BibitemOpen
  \bibfield  {author} {\bibinfo {author} {\bibfnamefont {A.}~\bibnamefont
  {Browaeys}}\ and\ \bibinfo {author} {\bibfnamefont {T.}~\bibnamefont
  {Lahaye}},\ }\bibfield  {title} {\bibinfo {title} {Many-body physics with
  individually controlled {Rydberg} atoms},\ }\href
  {https://doi.org/10.1038/s41567-019-0733-z} {\bibfield  {journal} {\bibinfo
  {journal} {Nat. Phys.}\ }\textbf {\bibinfo {volume} {16}},\ \bibinfo {pages}
  {132} (\bibinfo {year} {2020})}\BibitemShut {NoStop}%
\bibitem [{\citenamefont {Ebadi}\ \emph {et~al.}(2021)\citenamefont {Ebadi},
  \citenamefont {Wang}, \citenamefont {Levine}, \citenamefont {Keesling},
  \citenamefont {Semeghini}, \citenamefont {Omran}, \citenamefont {Bluvstein},
  \citenamefont {Samajdar}, \citenamefont {Pichler}, \citenamefont {Ho},
  \citenamefont {Choi}, \citenamefont {Sachdev}, \citenamefont {Greiner},
  \citenamefont {Vuletić},\ and\ \citenamefont {Lukin}}]{ebadi_quantum_2021}%
  \BibitemOpen
  \bibfield  {author} {\bibinfo {author} {\bibfnamefont {S.}~\bibnamefont
  {Ebadi}}, \bibinfo {author} {\bibfnamefont {T.~T.}\ \bibnamefont {Wang}},
  \bibinfo {author} {\bibfnamefont {H.}~\bibnamefont {Levine}}, \bibinfo
  {author} {\bibfnamefont {A.}~\bibnamefont {Keesling}}, \bibinfo {author}
  {\bibfnamefont {G.}~\bibnamefont {Semeghini}}, \bibinfo {author}
  {\bibfnamefont {A.}~\bibnamefont {Omran}}, \bibinfo {author} {\bibfnamefont
  {D.}~\bibnamefont {Bluvstein}}, \bibinfo {author} {\bibfnamefont
  {R.}~\bibnamefont {Samajdar}}, \bibinfo {author} {\bibfnamefont
  {H.}~\bibnamefont {Pichler}}, \bibinfo {author} {\bibfnamefont {W.~W.}\
  \bibnamefont {Ho}}, \bibinfo {author} {\bibfnamefont {S.}~\bibnamefont
  {Choi}}, \bibinfo {author} {\bibfnamefont {S.}~\bibnamefont {Sachdev}},
  \bibinfo {author} {\bibfnamefont {M.}~\bibnamefont {Greiner}}, \bibinfo
  {author} {\bibfnamefont {V.}~\bibnamefont {Vuletić}},\ and\ \bibinfo
  {author} {\bibfnamefont {M.~D.}\ \bibnamefont {Lukin}},\ }\bibfield  {title}
  {\bibinfo {title} {Quantum phases of matter on a 256-atom programmable
  quantum simulator},\ }\href {https://doi.org/10.1038/s41586-021-03582-4}
  {\bibfield  {journal} {\bibinfo  {journal} {Nature}\ }\textbf {\bibinfo
  {volume} {595}},\ \bibinfo {pages} {227} (\bibinfo {year}
  {2021})}\BibitemShut {NoStop}%
\bibitem [{\citenamefont {Ebadi}\ \emph {et~al.}(2022)\citenamefont {Ebadi},
  \citenamefont {Keesling}, \citenamefont {Cain}, \citenamefont {Wang},
  \citenamefont {Levine}, \citenamefont {Bluvstein}, \citenamefont {Semeghini},
  \citenamefont {Omran}, \citenamefont {Liu}, \citenamefont {Samajdar},
  \citenamefont {Luo}, \citenamefont {Nash}, \citenamefont {Gao}, \citenamefont
  {Barak}, \citenamefont {Farhi}, \citenamefont {Sachdev}, \citenamefont
  {Gemelke}, \citenamefont {Zhou}, \citenamefont {Choi}, \citenamefont
  {Pichler}, \citenamefont {Wang}, \citenamefont {Greiner}, \citenamefont
  {Vuletić},\ and\ \citenamefont {Lukin}}]{ebadi_optimization_2022}%
  \BibitemOpen
  \bibfield  {author} {\bibinfo {author} {\bibfnamefont {S.}~\bibnamefont
  {Ebadi}}, \bibinfo {author} {\bibfnamefont {A.}~\bibnamefont {Keesling}},
  \bibinfo {author} {\bibfnamefont {M.}~\bibnamefont {Cain}}, \bibinfo {author}
  {\bibfnamefont {T.~T.}\ \bibnamefont {Wang}}, \bibinfo {author}
  {\bibfnamefont {H.}~\bibnamefont {Levine}}, \bibinfo {author} {\bibfnamefont
  {D.}~\bibnamefont {Bluvstein}}, \bibinfo {author} {\bibfnamefont
  {G.}~\bibnamefont {Semeghini}}, \bibinfo {author} {\bibfnamefont
  {A.}~\bibnamefont {Omran}}, \bibinfo {author} {\bibfnamefont {J.-G.}\
  \bibnamefont {Liu}}, \bibinfo {author} {\bibfnamefont {R.}~\bibnamefont
  {Samajdar}}, \bibinfo {author} {\bibfnamefont {X.-Z.}\ \bibnamefont {Luo}},
  \bibinfo {author} {\bibfnamefont {B.}~\bibnamefont {Nash}}, \bibinfo {author}
  {\bibfnamefont {X.}~\bibnamefont {Gao}}, \bibinfo {author} {\bibfnamefont
  {B.}~\bibnamefont {Barak}}, \bibinfo {author} {\bibfnamefont
  {E.}~\bibnamefont {Farhi}}, \bibinfo {author} {\bibfnamefont
  {S.}~\bibnamefont {Sachdev}}, \bibinfo {author} {\bibfnamefont
  {N.}~\bibnamefont {Gemelke}}, \bibinfo {author} {\bibfnamefont
  {L.}~\bibnamefont {Zhou}}, \bibinfo {author} {\bibfnamefont {S.}~\bibnamefont
  {Choi}}, \bibinfo {author} {\bibfnamefont {H.}~\bibnamefont {Pichler}},
  \bibinfo {author} {\bibfnamefont {S.-T.}\ \bibnamefont {Wang}}, \bibinfo
  {author} {\bibfnamefont {M.}~\bibnamefont {Greiner}}, \bibinfo {author}
  {\bibfnamefont {V.}~\bibnamefont {Vuletić}},\ and\ \bibinfo {author}
  {\bibfnamefont {M.~D.}\ \bibnamefont {Lukin}},\ }\bibfield  {title} {\bibinfo
  {title} {Quantum optimization of maximum independent set using rydberg atom
  arrays},\ }\href {https://doi.org/10.1126/science.abo6587} {\bibfield
  {journal} {\bibinfo  {journal} {Science}\ }\textbf {\bibinfo {volume}
  {376}},\ \bibinfo {pages} {1209} (\bibinfo {year} {2022})},\ \Eprint
  {https://arxiv.org/abs/https://www.science.org/doi/pdf/10.1126/science.abo6587}
  {https://www.science.org/doi/pdf/10.1126/science.abo6587} \BibitemShut
  {NoStop}%
\bibitem [{\citenamefont {Knolle}\ and\ \citenamefont
  {Moessner}(2019)}]{KnolleMoessner_review}%
  \BibitemOpen
  \bibfield  {author} {\bibinfo {author} {\bibfnamefont {J.}~\bibnamefont
  {Knolle}}\ and\ \bibinfo {author} {\bibfnamefont {R.}~\bibnamefont
  {Moessner}},\ }\bibfield  {title} {\bibinfo {title} {A field guide to spin
  liquids},\ }\href
  {https://doi.org/https://doi.org/10.1146/annurev-conmatphys-031218-013401}
  {\bibfield  {journal} {\bibinfo  {journal} {Annual Review of Condensed Matter
  Physics}\ }\textbf {\bibinfo {volume} {10}},\ \bibinfo {pages} {451}
  (\bibinfo {year} {2019})}\BibitemShut {NoStop}%
\bibitem [{\citenamefont {Patil}(2023)}]{patil2023quantum}%
  \BibitemOpen
  \bibfield  {author} {\bibinfo {author} {\bibfnamefont {P.}~\bibnamefont
  {Patil}},\ }\href@noop {} {\bibinfo {title} {Quantum {M}onte {C}arlo
  simulations in the restricted hilbert space of {R}ydberg atom arrays}}
  (\bibinfo {year} {2023}),\ \Eprint {https://arxiv.org/abs/2309.00482}
  {arXiv:2309.00482 [cond-mat.str-el]} \BibitemShut {NoStop}%
\bibitem [{\citenamefont {Alet}\ \emph {et~al.}(2006)\citenamefont {Alet},
  \citenamefont {Ikhlef}, \citenamefont {Jacobsen}, \citenamefont {Misguich},\
  and\ \citenamefont {Pasquier}}]{Alet2006}%
  \BibitemOpen
  \bibfield  {author} {\bibinfo {author} {\bibfnamefont {F.}~\bibnamefont
  {Alet}}, \bibinfo {author} {\bibfnamefont {Y.}~\bibnamefont {Ikhlef}},
  \bibinfo {author} {\bibfnamefont {J.~L.}\ \bibnamefont {Jacobsen}}, \bibinfo
  {author} {\bibfnamefont {G.}~\bibnamefont {Misguich}},\ and\ \bibinfo
  {author} {\bibfnamefont {V.}~\bibnamefont {Pasquier}},\ }\bibfield  {title}
  {\bibinfo {title} {Classical dimers with aligning interactions on the square
  lattice},\ }\href {https://doi.org/10.1103/PhysRevE.74.041124} {\bibfield
  {journal} {\bibinfo  {journal} {Phys. Rev. E}\ }\textbf {\bibinfo {volume}
  {74}},\ \bibinfo {pages} {041124} (\bibinfo {year} {2006})}\BibitemShut
  {NoStop}%
\bibitem [{\citenamefont {Sahay}\ \emph {et~al.}(2023)\citenamefont {Sahay},
  \citenamefont {Vishwanath},\ and\ \citenamefont
  {Verresen}}]{sahay_quantum_2023}%
  \BibitemOpen
  \bibfield  {author} {\bibinfo {author} {\bibfnamefont {R.}~\bibnamefont
  {Sahay}}, \bibinfo {author} {\bibfnamefont {A.}~\bibnamefont {Vishwanath}},\
  and\ \bibinfo {author} {\bibfnamefont {R.}~\bibnamefont {Verresen}},\ }\href
  {https://doi.org/10.48550/arXiv.2211.01381} {\bibinfo {title} {Quantum {Spin}
  {Puddles} and {Lakes}: {NISQ}-{Era} {Spin} {Liquids} from {Non}-{Equilibrium}
  {Dynamics}}} (\bibinfo {year} {2023}),\ \bibinfo {note} {arXiv:2211.01381
  [cond-mat, physics:quant-ph]}\BibitemShut {NoStop}%
\bibitem [{\citenamefont {You}\ \emph {et~al.}(2007)\citenamefont {You},
  \citenamefont {Li},\ and\ \citenamefont {Gu}}]{you_fidelity_2007}%
  \BibitemOpen
  \bibfield  {author} {\bibinfo {author} {\bibfnamefont {W.-L.}\ \bibnamefont
  {You}}, \bibinfo {author} {\bibfnamefont {Y.-W.}\ \bibnamefont {Li}},\ and\
  \bibinfo {author} {\bibfnamefont {S.-J.}\ \bibnamefont {Gu}},\ }\bibfield
  {title} {\bibinfo {title} {Fidelity, dynamic structure factor, and
  susceptibility in critical phenomena},\ }\href
  {https://doi.org/10.1103/PhysRevE.76.022101} {\bibfield  {journal} {\bibinfo
  {journal} {Phys. Rev. E}\ }\textbf {\bibinfo {volume} {76}},\ \bibinfo
  {pages} {022101} (\bibinfo {year} {2007})}\BibitemShut {NoStop}%
\bibitem [{\citenamefont {Schwandt}\ \emph {et~al.}(2009)\citenamefont
  {Schwandt}, \citenamefont {Alet},\ and\ \citenamefont
  {Capponi}}]{schwandt_quantum_2009}%
  \BibitemOpen
  \bibfield  {author} {\bibinfo {author} {\bibfnamefont {D.}~\bibnamefont
  {Schwandt}}, \bibinfo {author} {\bibfnamefont {F.}~\bibnamefont {Alet}},\
  and\ \bibinfo {author} {\bibfnamefont {S.}~\bibnamefont {Capponi}},\
  }\bibfield  {title} {\bibinfo {title} {Quantum {Monte} {Carlo} {Simulations}
  of {Fidelity} at {Magnetic} {Quantum} {Phase} {Transitions}},\ }\href
  {https://doi.org/10.1103/PhysRevLett.103.170501} {\bibfield  {journal}
  {\bibinfo  {journal} {Phys. Rev. Lett.}\ }\textbf {\bibinfo {volume} {103}},\
  \bibinfo {pages} {170501} (\bibinfo {year} {2009})}\BibitemShut {NoStop}%
\bibitem [{\citenamefont {Albuquerque}\ \emph {et~al.}(2010)\citenamefont
  {Albuquerque}, \citenamefont {Alet}, \citenamefont {Sire},\ and\
  \citenamefont {Capponi}}]{albuquerque_quantum_2010}%
  \BibitemOpen
  \bibfield  {author} {\bibinfo {author} {\bibfnamefont {A.~F.}\ \bibnamefont
  {Albuquerque}}, \bibinfo {author} {\bibfnamefont {F.}~\bibnamefont {Alet}},
  \bibinfo {author} {\bibfnamefont {C.}~\bibnamefont {Sire}},\ and\ \bibinfo
  {author} {\bibfnamefont {S.}~\bibnamefont {Capponi}},\ }\bibfield  {title}
  {\bibinfo {title} {Quantum critical scaling of fidelity susceptibility},\
  }\href {https://doi.org/10.1103/PhysRevB.81.064418} {\bibfield  {journal}
  {\bibinfo  {journal} {Phys. Rev. B}\ }\textbf {\bibinfo {volume} {81}},\
  \bibinfo {pages} {064418} (\bibinfo {year} {2010})}\BibitemShut {NoStop}%
\bibitem [{\citenamefont {Wang}\ \emph {et~al.}(2015)\citenamefont {Wang},
  \citenamefont {Liu}, \citenamefont {Imriška}, \citenamefont {Ma},\ and\
  \citenamefont {Troyer}}]{wang_fidelity_2015}%
  \BibitemOpen
  \bibfield  {author} {\bibinfo {author} {\bibfnamefont {L.}~\bibnamefont
  {Wang}}, \bibinfo {author} {\bibfnamefont {Y.-H.}\ \bibnamefont {Liu}},
  \bibinfo {author} {\bibfnamefont {J.}~\bibnamefont {Imriška}}, \bibinfo
  {author} {\bibfnamefont {P.~N.}\ \bibnamefont {Ma}},\ and\ \bibinfo {author}
  {\bibfnamefont {M.}~\bibnamefont {Troyer}},\ }\bibfield  {title} {\bibinfo
  {title} {Fidelity susceptibility made simple: {A} unified quantum {Monte}
  {Carlo} approach},\ }\href {https://doi.org/10.1103/PhysRevX.5.031007}
  {\bibfield  {journal} {\bibinfo  {journal} {Phys. Rev. X}\ }\textbf {\bibinfo
  {volume} {5}},\ \bibinfo {pages} {031007} (\bibinfo {year} {2015})},\
  \bibinfo {note} {arXiv:1502.06969 [cond-mat, physics:physics,
  physics:quant-ph]}\BibitemShut {NoStop}%
\bibitem [{\citenamefont {Tarabunga}\ \emph {et~al.}(2022)\citenamefont
  {Tarabunga}, \citenamefont {Surace}, \citenamefont {Andreoni}, \citenamefont
  {Angelone},\ and\ \citenamefont {Dalmonte}}]{tarabunga_gauge-theoretic_2022}%
  \BibitemOpen
  \bibfield  {author} {\bibinfo {author} {\bibfnamefont {P.}~\bibnamefont
  {Tarabunga}}, \bibinfo {author} {\bibfnamefont {F.}~\bibnamefont {Surace}},
  \bibinfo {author} {\bibfnamefont {R.}~\bibnamefont {Andreoni}}, \bibinfo
  {author} {\bibfnamefont {A.}~\bibnamefont {Angelone}},\ and\ \bibinfo
  {author} {\bibfnamefont {M.}~\bibnamefont {Dalmonte}},\ }\bibfield  {title}
  {\bibinfo {title} {Gauge-{Theoretic} {Origin} of {Rydberg} {Quantum} {Spin}
  {Liquids}},\ }\href {https://doi.org/10.1103/PhysRevLett.129.195301}
  {\bibfield  {journal} {\bibinfo  {journal} {Phys. Rev. Lett.}\ }\textbf
  {\bibinfo {volume} {129}},\ \bibinfo {pages} {195301} (\bibinfo {year}
  {2022})}\BibitemShut {NoStop}%
\bibitem [{\citenamefont {Bricmont}\ and\ \citenamefont
  {Frölich}(1983)}]{bricmont_order_1983}%
  \BibitemOpen
  \bibfield  {author} {\bibinfo {author} {\bibfnamefont {J.}~\bibnamefont
  {Bricmont}}\ and\ \bibinfo {author} {\bibfnamefont {J.}~\bibnamefont
  {Frölich}},\ }\bibfield  {title} {\bibinfo {title} {An order parameter
  distinguishing between different phases of lattice gauge theories with matter
  fields},\ }\href {https://doi.org/10.1016/0370-2693(83)91171-1} {\bibfield
  {journal} {\bibinfo  {journal} {Physics Letters B}\ }\textbf {\bibinfo
  {volume} {122}},\ \bibinfo {pages} {73} (\bibinfo {year} {1983})}\BibitemShut
  {NoStop}%
\bibitem [{\citenamefont {Fredenhagen}\ and\ \citenamefont
  {Marcu}(1986)}]{fredenhagen_confinement_1986}%
  \BibitemOpen
  \bibfield  {author} {\bibinfo {author} {\bibfnamefont {K.}~\bibnamefont
  {Fredenhagen}}\ and\ \bibinfo {author} {\bibfnamefont {M.}~\bibnamefont
  {Marcu}},\ }\bibfield  {title} {\bibinfo {title} {Confinement criterion for
  {QCD} with dynamical quarks},\ }\href
  {https://doi.org/10.1103/PhysRevLett.56.223} {\bibfield  {journal} {\bibinfo
  {journal} {Phys. Rev. Lett.}\ }\textbf {\bibinfo {volume} {56}},\ \bibinfo
  {pages} {223} (\bibinfo {year} {1986})}\BibitemShut {NoStop}%
\bibitem [{\citenamefont {Fredenhagen}\ and\ \citenamefont
  {Marcu}(1988)}]{fredenhagen_dual_1988}%
  \BibitemOpen
  \bibfield  {author} {\bibinfo {author} {\bibfnamefont {K.}~\bibnamefont
  {Fredenhagen}}\ and\ \bibinfo {author} {\bibfnamefont {M.}~\bibnamefont
  {Marcu}},\ }\bibfield  {title} {\bibinfo {title} {Dual interpretation of
  order parameters for lattice gauge theories with matter fields},\ }\href
  {https://doi.org/10.1016/0920-5632(88)90124-7} {\bibfield  {journal}
  {\bibinfo  {journal} {Nuclear Physics B - Proceedings Supplements}\ }\textbf
  {\bibinfo {volume} {4}},\ \bibinfo {pages} {352} (\bibinfo {year}
  {1988})}\BibitemShut {NoStop}%
\bibitem [{\citenamefont {Gregor}\ \emph {et~al.}(2011)\citenamefont {Gregor},
  \citenamefont {Huse}, \citenamefont {Moessner},\ and\ \citenamefont
  {Sondhi}}]{gregor_diagnosing_2011}%
  \BibitemOpen
  \bibfield  {author} {\bibinfo {author} {\bibfnamefont {K.}~\bibnamefont
  {Gregor}}, \bibinfo {author} {\bibfnamefont {D.~A.}\ \bibnamefont {Huse}},
  \bibinfo {author} {\bibfnamefont {R.}~\bibnamefont {Moessner}},\ and\
  \bibinfo {author} {\bibfnamefont {S.~L.}\ \bibnamefont {Sondhi}},\ }\bibfield
   {title} {\bibinfo {title} {Diagnosing deconfinement and topological order},\
  }\href {https://doi.org/10.1088/1367-2630/13/2/025009} {\bibfield  {journal}
  {\bibinfo  {journal} {New J. Phys.}\ }\textbf {\bibinfo {volume} {13}},\
  \bibinfo {pages} {025009} (\bibinfo {year} {2011})}\BibitemShut {NoStop}%
\bibitem [{Note1()}]{Note1}%
  \BibitemOpen
  \bibinfo {note} {From the arguments we present below, the off-diagonal closed
  loop operator is also expected to be vanishingly small}\BibitemShut {NoStop}%
\bibitem [{Note2()}]{Note2}%
  \BibitemOpen
  \bibinfo {note} {Recall that the VBS consists of a localized subset of dimer
  configurations and can on small system sizes even have an overlap with the
  RVB state that is comparable to the one of the QSL with the RVB
  state.}\BibitemShut {Stop}%
\bibitem [{\citenamefont {Zeng}\ and\ \citenamefont
  {Elser}(1995)}]{ZengElser1995}%
  \BibitemOpen
  \bibfield  {author} {\bibinfo {author} {\bibfnamefont {C.}~\bibnamefont
  {Zeng}}\ and\ \bibinfo {author} {\bibfnamefont {V.}~\bibnamefont {Elser}},\
  }\bibfield  {title} {\bibinfo {title} {Quantum dimer calculations on the
  spin-1/2 kagom{\'e} heisenberg antiferromagnet},\ }\href
  {https://doi.org/10.1103/PhysRevB.51.8318} {\bibfield  {journal} {\bibinfo
  {journal} {Phys. Rev. B}\ }\textbf {\bibinfo {volume} {51}},\ \bibinfo
  {pages} {8318} (\bibinfo {year} {1995})}\BibitemShut {NoStop}%
\bibitem [{Sup()}]{Supplements}%
  \BibitemOpen
  \href@noop {} {\bibinfo  {journal} {See Supplemental Material at [URL will be
  inserted by publisher] for details on the quantum {M}onte {C}arlo
  simulations}\ }\BibitemShut {NoStop}%
\bibitem [{\citenamefont {Chen}\ \emph {et~al.}(2023)\citenamefont {Chen},
  \citenamefont {Bornet}, \citenamefont {Bintz}, \citenamefont {Emperauger},
  \citenamefont {Leclerc}, \citenamefont {Liu}, \citenamefont {Scholl},
  \citenamefont {Barredo}, \citenamefont {Hauschild}, \citenamefont
  {Chatterjee}, \citenamefont {Schuler}, \citenamefont {Läuchli},
  \citenamefont {Zaletel}, \citenamefont {Lahaye}, \citenamefont {Yao},\ and\
  \citenamefont {Browaeys}}]{Chen2023}%
  \BibitemOpen
\bibfield  {journal} {  }\bibfield  {author} {\bibinfo {author} {\bibfnamefont
  {C.}~\bibnamefont {Chen}}, \bibinfo {author} {\bibfnamefont {G.}~\bibnamefont
  {Bornet}}, \bibinfo {author} {\bibfnamefont {M.}~\bibnamefont {Bintz}},
  \bibinfo {author} {\bibfnamefont {G.}~\bibnamefont {Emperauger}}, \bibinfo
  {author} {\bibfnamefont {L.}~\bibnamefont {Leclerc}}, \bibinfo {author}
  {\bibfnamefont {V.~S.}\ \bibnamefont {Liu}}, \bibinfo {author} {\bibfnamefont
  {P.}~\bibnamefont {Scholl}}, \bibinfo {author} {\bibfnamefont
  {D.}~\bibnamefont {Barredo}}, \bibinfo {author} {\bibfnamefont
  {J.}~\bibnamefont {Hauschild}}, \bibinfo {author} {\bibfnamefont
  {S.}~\bibnamefont {Chatterjee}}, \bibinfo {author} {\bibfnamefont
  {M.}~\bibnamefont {Schuler}}, \bibinfo {author} {\bibfnamefont {A.~M.}\
  \bibnamefont {Läuchli}}, \bibinfo {author} {\bibfnamefont {M.~P.}\
  \bibnamefont {Zaletel}}, \bibinfo {author} {\bibfnamefont {T.}~\bibnamefont
  {Lahaye}}, \bibinfo {author} {\bibfnamefont {N.~Y.}\ \bibnamefont {Yao}},\
  and\ \bibinfo {author} {\bibfnamefont {A.}~\bibnamefont {Browaeys}},\
  }\bibfield  {title} {\bibinfo {title} {Continuous symmetry breaking in a
  two-dimensional {R}ydberg array},\ }\href
  {https://doi.org/10.1038/s41586-023-05859-2} {\bibfield  {journal} {\bibinfo
  {journal} {Nature}\ } (\bibinfo {year} {2023})}\BibitemShut {NoStop}%
\bibitem [{\citenamefont {Sbierski}\ \emph {et~al.}(2024)\citenamefont
  {Sbierski}, \citenamefont {Bintz}, \citenamefont {Chatterjee}, \citenamefont
  {Schuler}, \citenamefont {Yao},\ and\ \citenamefont {Pollet}}]{Sbierski2024}%
  \BibitemOpen
  \bibfield  {author} {\bibinfo {author} {\bibfnamefont {B.}~\bibnamefont
  {Sbierski}}, \bibinfo {author} {\bibfnamefont {M.}~\bibnamefont {Bintz}},
  \bibinfo {author} {\bibfnamefont {S.}~\bibnamefont {Chatterjee}}, \bibinfo
  {author} {\bibfnamefont {M.}~\bibnamefont {Schuler}}, \bibinfo {author}
  {\bibfnamefont {N.~Y.}\ \bibnamefont {Yao}},\ and\ \bibinfo {author}
  {\bibfnamefont {L.}~\bibnamefont {Pollet}},\ }\bibfield  {title} {\bibinfo
  {title} {Magnetism in the two-dimensional dipolar {XY} model},\ }\href
  {https://doi.org/10.1103/PhysRevB.109.144411} {\bibfield  {journal} {\bibinfo
   {journal} {Phys. Rev. B}\ }\textbf {\bibinfo {volume} {109}},\ \bibinfo
  {pages} {144411} (\bibinfo {year} {2024})}\BibitemShut {NoStop}%
\end{thebibliography}%

\newpage~
\newpage~

\setcounter{equation}{0}
\setcounter{figure}{0}
\setcounter{table}{0}
\setcounter{section}{0}
\setcounter{page}{1}
\renewcommand{\theequation}{S\arabic{equation}}
\renewcommand{\thefigure}{S\arabic{figure}}
\renewcommand{\thetable}{S\Roman{table}}

\onecolumngrid

\begin{center} \huge{Supplemental Material: The renormalized classical spin liquid on the ruby lattice\\}
\end{center}


\vspace*{2cm}

In the sections below we present additional results to support our claims in the main paper.

\section{Spectra}
\label{app:spectra}

\begin{figure}[h]
    \centering
    \includegraphics[width=0.8\textwidth]{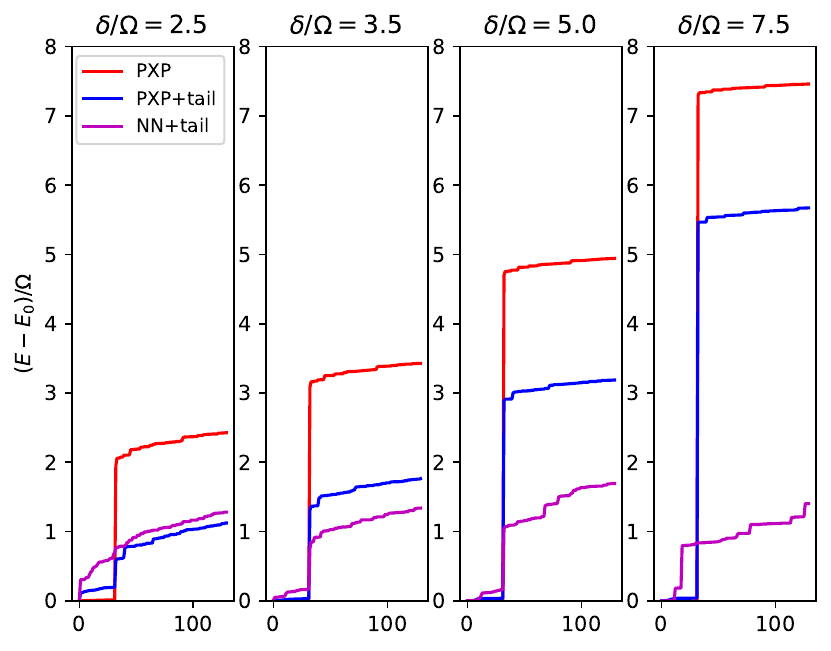}   
    \caption{ 
    Spectra of the lowest 130 eigenvalues obtained by exact diagonalization for a ruby lattice with periodic boundaries of size $L=2$ for various values of $\delta$ and different potentials: (i) {\it PXP} is the blockade model with infinite repulsion inside the blockade radius $R_b=2.4a$, (ii) {\it PXP}+tail is the {\it PXP} model augmented with Van der Waals tails outside the blockade radius, (iii) NN+tail has a blockade radius $R_b=1.00001$ and Van der Waals interactions outside it. Whenever the red curve is not visible, the blue curve is on top of it. Lines are a guide to the eye. }
    \label{fig:spectra}
\end{figure}

We present in Fig.~\ref{fig:spectra} spectra for small system sizes $L \times L =2 \times 2$ obtained with full diagonalization and Lanczos methods. It contains instructive information to appreciate the entropy density plots and the separation of electric and magnetic scales from the spin-liquid point of view. \\

For all shown values of $\delta$ in the {\it PXP} model there is a low-energy manifold consisting of $2^5=32$ states, as expected for 4 hexagons. All states in this manifold are in the electric ground state, with expectation values of the Gauss law (corresponding to the product of 4 Pauli $Z$ operators on the 4 sites closest to one Kagome vertex) on all vertices close to 1. The magnetic (plaquette) degrees of freedom are defined as the product of an X-string around each single hexagon as in Ref.~\cite{semeghini_probing_2021}, which connects to the "arrow representation"\cite{ZengElser1995,misguich_2002} in the full dimer limit. Those plaquette degrees of freedom are of order 1 in the ground state, and in the low energy manifold there are 4 states with plaquette values close to +1. Such a spectrum is expected for a toric code model, with strong anisotropy between the electric and magnetic degrees of freedom. Note that a splitting, exponentially small in the system size, of the 4-fold ground state due to finite size effects is routinely seen in exact diagonalization studies of quantum spin liquids. Furthermore, the low energy manifold is seen to scale as $\sim \Omega^6/\delta^5$. For $\delta = 2$ the gap is $\sim 0.005$, implying that QMC requires inverse temperatures $\beta \gg 200$ and a precision on the energy per site on the 4th digit, which is out of reach. In the low energy manifold, we see no difference between a QSL and a VBS because of the too small system size: The difference between the two is that the QSL is thought of as an equal-weight, positive phase superposition of all dimer states, whereas the VBS consists of a localized subset of some of the dimer states. 
Interestingly, the lowest state of the manifold with an electric excitation (state number 32 in the plots and with an energy $\sim \delta$ above the ground state) still has 4 well-defined magnetic excitations with values of order $1/4$, and all 4 are positive. In exact diagonalization, this state is found below the rest of the manifold by an amount proportional to the magnetic interaction scale. In the DMRG simulations of Ref.~\cite{semeghini_probing_2021}, this state was found to be highly populated following a realistic ramp protocol (see Fig. S15 of that paper). \\

In the {\it PXP}+tail model, we add Van der Waals tail to the {\it PXP} model. The spectra are very similar to the one of the pure {\it PXP} model, with the exception of $\delta=2.5$ where the ground state is substantially lower than the other states in the low energy manifold. This is interpreted as a shift in the location of the phase transition between the trivial and the QSL phases to larger values of $\delta$. Otherwise, the tails cannot have a strong influence (because the system size is too small to host a VBS phase).  \\

In order to model the Van der Waals interactions, we approximate the potential on nearest neighbors (taking a value $\sim 200$) with an infinite value, which reduces the size of the Hilbert dimension to just $65536$ for $L=2$ without taking symmetries into account. The spectrum for $\delta=2.5$ is quite different from the one of its {\it PXP} cousins, and clearly indicative of a trivial phase. For values of $\delta$ in the range 3-7 (this range corresponds to values of the VdW potential between the second and third nearest neighbors inside the blockade radius), the spectrum still shows a two-scale behavior with 32 states in the low-energy sector but with additional hybridization between magnetic and electric degrees of freedom. The fate of the ground state cannot be inferred from exact diagonalization for such small system sizes. DMRG studies on larger cylinder geometries find a QSL over a very narrow region around $\delta \approx 3.5$ when the interactions are truncated below $A=6a$. Furthermore, based on the structure seen inside the manifold of the 32 lowest states for $\delta=3.5$ and $\delta=5$ in Fig.~\ref{fig:spectra}, we expect that the entropy of the classical spin liquid will not be pinned to $\log(2)/6$ in the thermodynamic limit but rather be $\delta-$dependent, as was observed in the QMC simulations. For values of $\delta > 6.5$, the spectrum is manifestly deviating from the one of the {\it PXP} model, from which the formation of a robust crystalline ground state can clearly be expected.

\section{Supporting results from quantum Monte Carlo simulations }
\label{app:FSS_QMC}

\subsection{Diagonal string parity loops}

\begin{figure}
    \centering
    \includegraphics[width=0.9\textwidth]{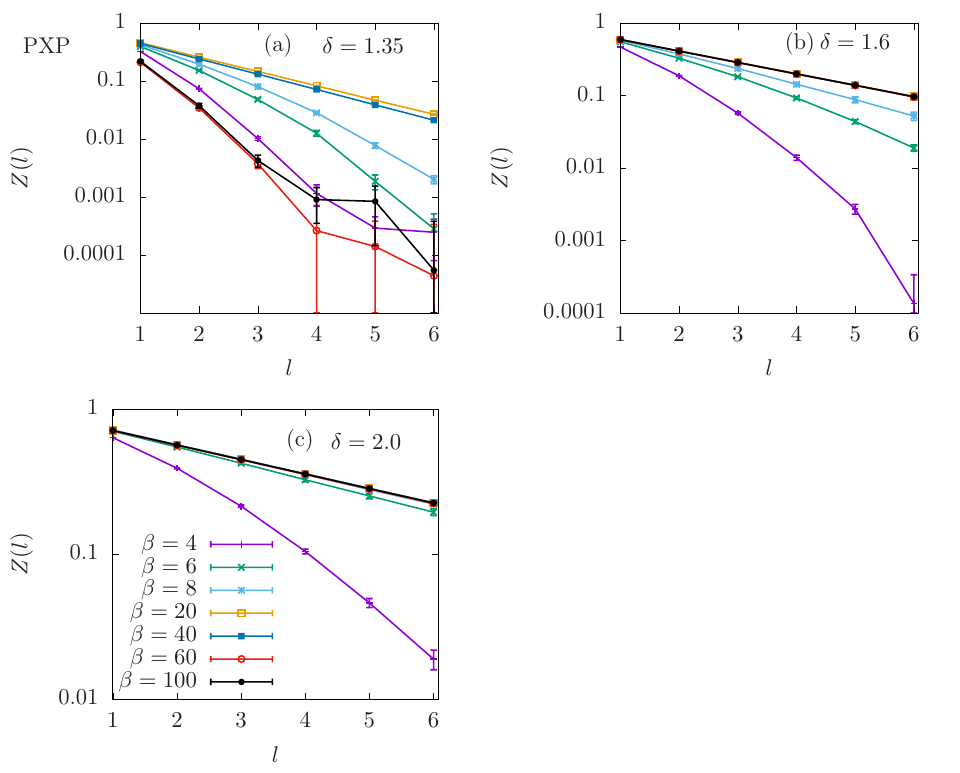}  
    \caption{ Length $l$ dependence of the diagonal loop operator from $l=1$ to $6$ at various temperatures for PXP model, in the cases of $\delta=1.35$, $1.6$ and $2.0$.  
    }
    \label{fig:PXP_loop}
\end{figure}

In Fig.~\ref{fig:PXP_loop} we show additional data for the closed-loop diagonal string parity operator $\mathcal{Z}(l)$ for the {\it PXP} model. For (a) $\delta=1.35$, $\mathcal{Z}(l)$ scales according to a perimeter law for $\beta=20-40$. At lower temperature, the CSL yields to the trivial paramagnet, causing a much faster reduction of $\mathcal{Z}(l)$ with increasing $l$. For (b) $\delta = 1.6$ and (c) $\delta = 2$, perimeter scaling is seen for $\beta \ge 20$ and $\beta \ge 8$ respectively down to the lowest temperature we could simulate, $\beta=100$. This is indicative of the broad CSL in which the system is in its ground state with respect to the electric degrees of freedom but incoherent with respect to the magnetic degrees the formation. The possible formation of a VBS is due to a resonance within the magnetic sector but can only happen at much lower temperatures, and plays no role for the physics described in this paper. \\

\begin{figure}
    \centering
    \includegraphics[width=0.9\textwidth]{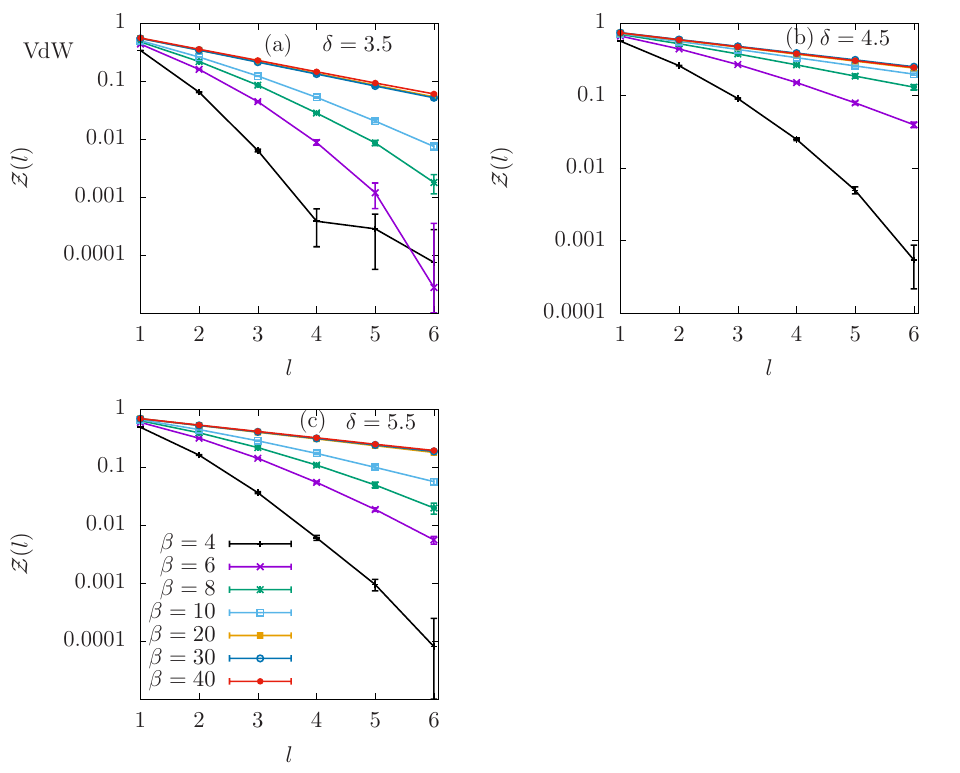}   
    \caption{ Same as Fig.~\ref{fig:PXP_loop}  for the model with  realistic VdW interactions truncated at a distance $A=5a$. 
      }
    \label{fig:VdW_loop}
\end{figure}

In Fig.~\ref{fig:VdW_loop} we show additional data for the closed-loop diagonal string parity operator $\mathcal{Z}(l)$ for the VdW model with truncation at distance $A=5a$, which is chosen to speed up the computation time without affecting the results. Note that for nearly all temperatures for which we show data the longer-range tails of the potential would be cut off by temperature. Recall that the cases (a) $\delta=3.5$ and (b) $\delta=4.5$ are the limits of the range over which the onset of a dynamically prepared QSL was seen~\cite{semeghini_probing_2021}. In both cases, QMC finds the formation of a CSL at low enough temperature ($\beta \ge 20)$ with uniform perimeter scaling of $\mathcal{Z}(l)$. The assumption that the electric degrees of freedom are in equilibrium at low but finite temperature is however not compatible with the two-scale interpretation given to the experimental data in Figs. 2C and 4H of Ref.~\cite{semeghini_probing_2021}. Attempting to match the order of magnitude of $Z(l), l=2 \ldots 4$, between experiment and QMC yields temperature estimates in the range $\beta = 5-8$. We stress again that these are order of magnitude estimates under the assumption of full equilibrium for the electric degrees of freedom, which is a very strong assumption for a many-body operator like $\mathcal{Z}(l)$. 

The case of $\delta=5.5$ (Fig.~\ref{fig:PXP_loop}(c)) is added for completeness. It is the parameter regime where the CSL has an entropy density $S/N \approx \log(2)/6$, but its ground state is known to be already deep in the crystalline phase even for finite truncation of the VdW potential.  \\

\subsection{Finite size scaling of local thermodynamic observables}

\begin{figure}
    \centering
    \includegraphics[width=0.75\textwidth]{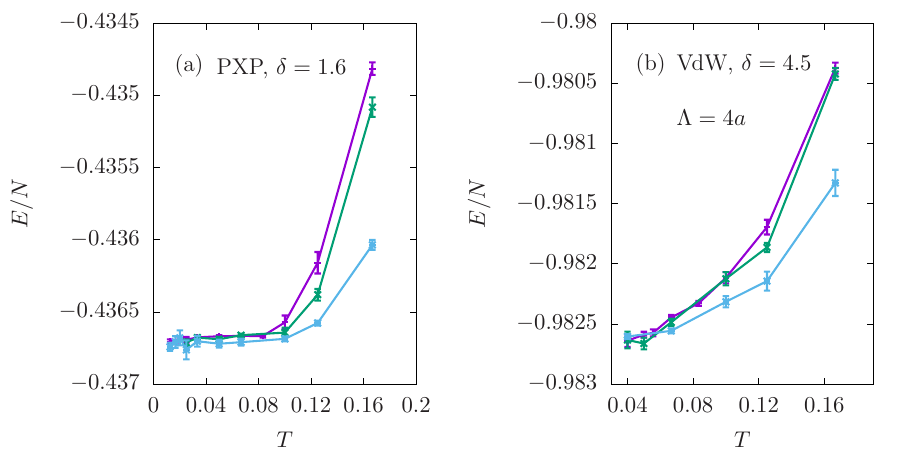}   
    \caption{ Temperature dependence of the energy density for $L=4,8$ and $12$, of the {\it PXP} model (a) and the Van der Waals interaction case (b). In the CSL the energy densities show no variation with system size up to the 4th digit.
    } 
    \label{fig:energy}
\end{figure}

The QMC has found a classical spin liquid characterized by electric degrees of freedom strongly localized on the vertices. From spectral and perturbative considerations, we know that the (independent) magnetic excitations would be strongly localized on the plaquettes. This picture is particularly strong in the {\it PXP} model and we would expect very small finite size effects in the CSL phase, which also carries over to the CSL phase of in the VdW model. In Fig.~\ref{fig:energy} we show that this is indeed the case. Of course, big system sizes are still welcome to distinguish area law from perimeter law scaling. Stronger finite size effects are expected at higher temperature, and for the formation of the VBS in the ground state which is of no concern to the discussion in this paper.

\end{document}